\documentclass[preprint,aps,showpacs]{revtex4}
\usepackage{graphicx}

\usepackage{amsmath}
\usepackage{amsxtra}
\usepackage{amstext}
\usepackage{amssymb}
\usepackage{latexsym}
\usepackage{dsfont}

\newcommand{\be}{\begin{equation}}
\newcommand{\ee}{\end{equation}}
\newcommand{\bn}{\begin{eqnarray}}
\newcommand{\en}{\end{eqnarray}}

\begin{document}

\title{Properties of twice four families of quarks and
leptons, of scalars and gauge fields \\ as predicted by the {\it spin-charge-family} theory}

\author{Norma Susana Manko\v c  
Bor\v stnik}
\affiliation{Department of Physics, FMF, University of Ljubljana,
Jadranska 19, Ljubljana, 1000}

%\maketitle

\begin{abstract}

The {\em  spin-charge-family} theory, proposed by the author~\cite{norma,pikanorma} as a possible 
new way to explain 
the assumptions of the {\it standard model}, predicts at the low energy regime two decoupled groups of 
four families of quarks and leptons. In two successive breaks the massless families, first the 
group of four and at the second break the rest four families, gain nonzero mass matrices. The 
families are identical with respect to the charges and spin. \\
There are two kinds of fields in this theory, which manifest at low energies as the gauge 
vector and scalar fields: the fields which couple to the charges and spin, and the fields 
which couple to the family quantum numbers. In loop corrections  to the tree level mass matrices 
both kinds start to contribute coherently. 
The fourth family of the lower group of four families is predicted to be possibly observed 
at the LHC and the stable of the higher four families -- the fifth family -- is the candidate 
to constitute the dark matter. 
Properties of the families of quarks and leptons and of the scalar and gauge fields, before and 
after each of the two  successive breaks, bringing masses to fermions and to boson fields are 
analysed and relations among coherent contributions of the loop corrections to fermion properties 
discussed, including the one which enables the existence of the Majorana neutrinos. 
The relation of the scalar fields and mass matrices following from the 
{\em  spin-charge-family} theory to the {\it standard model} Yukawa couplings and Higgs is discussed. 
Although effectively the scalar fields manifest as Higgs and Yukawas, measurements of the 
scalar fields do not coincide with the measurement of the Higgs. 
Also the relation to the attempts which try to explain the family quantum number with the $SU(3)$ 
flavour groups and the Yukawas as scalar dynamical fields in the "bi-fundamental" representation 
of the $SU(3)$ flavour groups is presented.  More detailed analyses of mass matrices 
with numerical results are in preparation~\cite{albinonorma}. 

\end{abstract}


\maketitle

\section{Introduction } 
\label{introduction}

The {\it standard model } offered more than 35 years ago  an elegant 
next step in understanding the origin of fermions and bosons. It is built on 
several assumptions leaving many open questions to be answered in the next step of  the 
theoretical interpretations. 
A lot of proofs and calculations have been done which support the {\it standard model}.

The measurements so far offer no sign which would help to make the next step below the 
{\it standard model}.

To explain  the assumptions, which are the building blocs of the {\it standard model}  
and to make successful new step beyond it, any new proposal, model, theory must  in my opinion 
at least  
i.) explain the origin of families and their mass matrices, 
 predict the number of families, and accordingly explain the Yukawa couplings, mixing matrices and masses 
 of family members,
ii.) explain the origin of  the {\it standard model} scalar  fields (the Higgs) and
their connections with  masses and Yukawa couplings, and 
iii.) explain the origin of the  dark matter. 
These three open 
questions are to my understanding so tightly connected that they call for common explanation.

Without a theory which is offering the answers to at least the above questions  the suggestions for what 
and how should experiments search for new events can hardly be successful. 

The {\it theory unifying spin and charges and predicting 
families}~\cite{norma,pikanorma,Portoroz03,gmdn,gn}, to be called the {\it spin-charge-family}theory,
seems  promising in answering these, and several other questions, which the {\it standard model} 
leaves unanswered. 

%Let me in what follows explain the achievements of the {\it spin-charge-family} theory by looking at
%the assumptions of the {\it standard model} through this theory. 

The {\it spin-charge-family} theory assumes in 
$d= (1+ (d-1))$, $d=14$ (or larger),  a  simple starting action for spinors and the gauge fields: Spinors 
carry only two kinds of the spins (no charges), namely the one postulated by Dirac 80 years ago and the second 
kind proposed by the author of this paper. There is no  third kind of a spin. Spinors  
interact  with only  vielbeins and the two kinds of the corresponding 
spin connection fields.

After the   
breaks of the starting symmetry, leading to the low energy regime, the simple starting 
action~(Eq.(\ref{wholeaction})) manifests  two decoupled groups of four families of quarks and leptons, 
with only the left handed (with respect to $d=\,(1+3)$) members of each family carrying the weak 
charge while the right handed ones are weak chargeless. The fourth family is 
predicted~\cite{pikanorma,gmdn} to be possibly observed at the LHC or at somewhat higher energies, 
while the stable fifth 
family members, forming neutral (with respect to the colour and electromagnetic 
charge) baryons and the fifth family neutrinos are predicted to explain the origin of the 
dark matter~\cite{gn}. 
 
 The spin connections,  associated with the two kinds of spins, together with vielbeins, 
 all behaving as scalar fields with  respect to  $d=\,(1+3)$,  are with their vacuum expectation 
 values at the two $SU(2)$ breaks  responsible for the nonzero mass matrices of fermions 
 and also for the masses of the gauge fields. 
 The spin connections with the indices of  vector fields with  respect to  $d=\,(1+3)$,  
 manifest after the break of symmetries  as the known gauge fields.

Although the properties of the scalar fields, that is their vacuum expectation values, coupling 
constants and masses, 
can not be calculated without the detailed knowledge of the mechanism of breaking the symmetries, and 
have been so far only roughly estimated, 
yet one can see, assuming  breaks which lead to observable phenomena at the low energy regime, 
how the properties of the scalar fields determine the fermion mass matrices,  
manifesting  effectively as the {\it standard model} Higgs and its Yukawa couplings.

According to the {\it spin-charge-family} theory the break of the starting symmetry is caused by  nonzero 
expectation values of  both kinds of  the spin connection fields together with the vielbeins, which all 
are scalars with respect to 
%the remaining 
$SO(1,3)$ symmetry. At the  symmetry  of $SO(1,7) \times U(1)_{II} 
\times SU(3)$ there are ($2^{\frac{1+7}{2}-1}$) massless left handed (with respect to $SO(1,7)$) 
families~\footnote{We studied  in the 
refs.~\cite{dhn,DHN}  a toy model, in which fermions, gauge and scalar fields live in $M^{1+5}$, 
which is  broken into $M^{1+3}\times$ an infinite disc. One can find several conditions, under 
which only left handed family members stay massless.}, which stay massless until  the symmetry 
$SO(1,3) \times SU(2)_{I} \times SU(2)_{II} 
\times U(1)_{II} \times SU(3)$ breaks. 

Four of the eight families are doublets with respect to the generators (Eqs.~(\ref{so13},\ref{so4},% 
\ref{so6}))  $\vec{\tilde{\tau}}^2$ and with respect to $\vec{\tilde{N}}_{R}$, while they are 
singlets with respect to $\vec{\tilde{\tau}}^1$ and $\vec{\tilde{N}}_{L}$. The other four 
families are singlets with respect to the first two kinds of generators and doublets with
respect to the second two kinds ($\vec{\tilde{\tau}}^1$ and $\vec{\tilde{N}}_{L}$) of generators. 

Each member of these eight massless families carries before the two successive breaks,   
from  $SO(1,3) \times SU(2)_{I} \times SU(2)_{II} \times U(1)_{II} \times SU(3)$ (first to 
 $SO(1,3) \times SU(2)_{I} \times U(1)_{I} \times SU(3)$ and then to  $SO(1,3)  
\times  U(1) \times SU(3)$) the quantum numbers of the two $SU(2)$, one $U(1)$ and
one $SU(3)$ charges and the quantum numbers of the subgroups to which each of the four families belong 
with respect to $\vec{\tilde{\tau}}^{(2,1)}$ and $\vec{\tilde{N}}_{(R,L)}$. 

Analysing properties of each family member with respect to the quantum numbers of the charge and 
spin subgroups, $SO(1,3), \,SU(2)_I,\, SU(2)_{II},\, U(1)_{II}$ and $SU(3)$, we find that each family 
includes left handed (with respect to $SO(1,3)$) weak $SU(2)_{I}$ charged quarks and leptons and 
right handed (again with respect to $SO(1,3)$) weak $SU(2)_{I}$ chargeless quarks and leptons. 
While the right handed members (with respect to $SO(1,3)$) are doublets with respect to $SU(2)_{II}$, 
the left handed fermions are singlets with respect to $SU(2)_{II}$.

Each member of the eight massless families of quarks and leptons  carries, if right handed, 
the $SU(2)_{II}$ charge,  in addition to the family quantum 
numbers and the quantum numbers of the {\em standard model}.
This $SU(2)_{II}$ charge determines, together with the $U(1)_{II}$ charge, after the first of the two 
breaks the {\em standard model} hyper charge and the fermion quantum number ($-\frac{1}{2}$ for leptons and  
$\frac{1}{6}$ for quarks).

Each break is triggered by  a particular superposition of one or  both   kinds of the spin connection  and 
the vielbeins. %To be in agreement in the low energy regime with the {\it standard model} assumptions supported  
%by the experimental data,  the gauge fields which are scalars with respect to $SO(1,3)$ 
%and have appropriate symmetries  are assumed to contribute. 
%
In the break of $SU(2)_{I} \times SU(2)_{II} \times U(1)_{II} $ to  
$SU(2)_{I} \times U(1)_{I}$ the scalar (with respect to $d=\, (1+3)$) fields originating 
in vielbeins and in  the second kind of  spin connection fields belonging to 
a triplet with respect to the $SU(2)_{II}$ symmetry in the $\tilde{S}^{ab}$ sector (with the generators  
$\tilde{\tau}^{2i}= c^{2i}{}_{ab}\tilde{S}^{ab}$, $\{\tilde{\tau}^{2i},\tilde{\tau}^{2j}\}_{-} = 
\varepsilon^{ijk} \tilde{\tau}^{2k}$) and with respect to one of the two $SU(2)$ from $SO(1,3)$ again in 
the  $\tilde{S}^{ab}$ sector (with the generators $\tilde{N}_{R}^{i}= c^{Ri}{}_{ab}\tilde{S}^{ab}$, 
$\{\tilde{N}_{R}^{i},\tilde{N}_{R}^{j}\}_{-} = \varepsilon^{ijk} \tilde{N}_{R}^{k}$) gain nonzero vacuum 
expectation values. 

The upper four families, which are doublets with respect to these 
two $SU(2)$ groups, become massive and so does the $SU(2)_{II}$ gauge vector field (in the adjoint 
representation of $SU(2)_{II}$ in the $S^{ab}$ sector with $\tau^{2i}=c^{2i}{}_{ab} S^{ab} $, 
$\{\tau^{2i},\tau^{2j}\}_{-}= \varepsilon^{ijk} \tau^{2k}$). 
The lower four families, which are singlets with respect to $\tilde{\tau}^{2i}$ and $\tilde{N}_{R}^{i}$,
the  two $SU(2)$ subgroups  in the $\tilde{S}^{ab}$ sector, stay massless. 

This is assumed to happen below the energy scale of $10^{13}$ GeV, that is below the 
unification scale of all the three 
charges, and also pretty much above the electroweak break.
 
The lower four families become massive at the electroweak break, when $SU(2)_{I} 
\times U(1)_{I}$ breaks into $U(1)$. To this break the vielbeins and the scalar part of both kinds  
of the spin connection fields contribute, those which are triplets with respect to the two remaining 
invariant $SU(2)$ subgroups in the $\tilde{S}^{ab}$ sector, $\tilde{\tau}^{1i}$ ($\tilde{\tau}^{1i}= 
c^{1i}{}_{ab}\tilde{S}^{ab}$) and $\tilde{N}_{L}^{i}$ ($\tilde{N}_{L}^{i}= c^{Li}{}_{ab}\tilde{S}^{ab}$), 
as well as the scalar gauge fields of $Q,Q'$ and $Y'$ (all expressible with $S^{ab}$). In this break 
also the $SU(2)_{I}$ weak gauge vector field becomes massive.

Although the estimations of the properties of families done so far are very approximate~\cite{gmdn,gn},
yet the predictions  give a hope that the starting assumptions  of the {\it spin-charge-family} theory 
are the right ones: 
 
\noindent i. Both existing  Clifford algebra operators determine 
properties of fermions. The Dirac $\gamma^a$'s  manifest in the low energy regime  
the spin and all the charges of fermions (like in the Kaluza-Klein  
theories~\footnote{The Kaluza-Klein[like] theories have difficulties with 
(almost) masslessness of the spinor fields at the low energy regime. In the refs.~\cite{DHN,dhn} we 
are proposing possible solutions to these kind of difficulties.}). 
The second kind of the spin, 
forming the equivalent representations with 
respect to the Dirac one,  manifests  families of fermions. 

\noindent ii. Fermions carrying only 
the corresponding two kinds of the spin (no charges) interact with the gravitational 
fields -- the vielbeins and (the two kinds of) the spin connections. The spin connections originating 
in  the Dirac's gammas ($\gamma^a$'s) manifest at the low energy regime  the known gauge fields.
The spin connections originating in the second 
kind of gammas ($\tilde{\gamma}^a$'s) are responsible, together with the vielbeins and  the spin connections 
of the first kind, for the masses of fermions and
gauge fields. 

\noindent iii. The assumed starting 
action for spinors and  gauge fields in $d$-dimensional space is simple: In $d$-dimensional space 
all the fermions are massless and interact with the corresponding gauge fields of the Poincar\' e group 
and the second kind of the spin connections, the corresponding Lagrange densities for the gauge fields 
are linear in the two Riemann tensors.

The project to come from the starting action through breaks of 
symmetries to the effective action at low (measurable) energy regime is very demanding. Although 
one easily sees that a part of the starting action manifests, after the breaks of symmetries, at the tree 
level the mass matrices of the families and that a part of the vielbeins 
together with the two kinds of the spin connection fields manifest as scalar fields, 
yet  several proofs are still needed besides those done so far~\cite{DHN,dhn} 
to guarantee that the {\it spin-charge-family} theory does lead to the measured 
effective action of the {\it standard model}.  Additional proofs and very demanding calculations  
in addition to  rough estimations~\cite{pikanorma,gmdn,gn} done so far are needed to show that predictions 
agree also with the measured values of  masses and mixing matrices of the so  far observed fermion families, 
explaining where do large differences among  masses of quarks and leptons, as 
well as among their mixing matrices originate.  Although it is not difficult to see that the 
differences must origin in charges which family members carry and in the fact that the closer are the  mass 
matrices  (in the massless basis) to the democratic ones, the smaller are the lower three masses in 
comparison with the fourth one.

Let us point out that in the {\em spin-charge-family} theory the scalar  (with respect to $(1+3)$) 
spin connection fields, originating in the Dirac kind of spin, 
couple only to the charges and spin, contributing on the tree level equally to all the families, 
distinguishing only among the members of one family (among the u-quark, d-quark, neutrino and 
electron, the left and right handed), the other scalar spin connection fields, originating in the 
second kind of spin, couple only to the family quantum numbers.  
Both kinds start to contribute coherently only beyond the tree level  and 
a detailed study should manifest the drastic differences in properties of quarks and leptons: 
in their masses and mixing matrices~\cite{norma,Portoroz03}.
It is a hope that the  loop corrections will help to understand  the differences in properties of 
fermions, with neutrinos included and the calculations will show to which extent  are 
the Majorana terms responsible for the great difference in the properties of neutrinos and the 
rest of the family members.

In this work the mass matrices of the two groups of four families,  
the two groups of the scalar fields giving masses to the two groups of four families and to the 
gauge fields to which they couple, and the gauge fields  are studied and their properties discussed, as 
they follow from the {\it spin-charge-family} theory. 
Many an assumption, presented above, allowed by the {\it spin-charge-family} theory, 
is made in order that the low energy manifestation of the theory agrees with the observed phenomena, but
not (yet) proved that it  follows dynamically from the theory.

The main purpose of this paper is  to manifest that there is a chance 
that the properties of the observed three families -- as members of the four coupled families --
naturally follow equivalently for quarks and leptons from the {\it spin-charge-family} theory 
when going beyond the tree level, although on 
the tree level the mass matrices of leptons and quarks are strongly related.  
A possible explanation is made why  the observed family members 
differ so much in their properties. 
It is also explained why does the  {\it spin-charge-family} theory predict 
two stable families, and why and how much do the 
fifth family hadrons  -- offering the explanation for the existence of the dark matter -- 
differ in their properties from the properties of the lower group of 
four families ones.

In the refs.~\cite{pikanorma,gmdn} we studied the properties of the lower four 
families under the assumption that the loop corrections  would  not change much the 
symmetries of the mass matrices of the family members, as they follow from the 
{\em spin-charge-family} theory on the 
tree level, but would take care of differences in properties among members. 
Relaxing  strong connections between the 
mass matrices of the $u$-quark and neutrino and the $d$-quark and electron, we were able 
to predict some properties of the fourth family members and their mixing matrix elements 
with the three so far measured families.
In this paper the %justification for the  
relaxation is discussed. 
The concrete evaluations of the properties of the mass matrices beyond the 
tree level are  in progress. First steps are done in the contribution with 
A. Hern\'andez-Galeana~\cite{albinonorma}, while more detailed analyses of the mass matrices 
with numerical results are in preparation.

The {\it standard model} is presented as a low energy effective theory 
of the {\it spin-charge-family} theory. Also  some  attempts in the literature to understand 
families as the $SU(3)$ flavour extension of the {\it standard model} are commented.

\section{The {\em spin-charge-family} theory from the starting action to the  {\em standard model} action}
\label{actiontosm}

Let as add in this section to the introduction into the {\it spin-charge-family} theory, made in the previous  
section, the mathematical part. The theory assumes that the spinor carries 
in $d(=(1 + 13))$-dimensional 
space  two kinds of the spin, no charges~\cite{norma}:
$\,\,$ 
{\bf i.} The Dirac spin, described by $\gamma^a$'s, defines the  spinor representations in $d=(1+ 13)$, and 
correspondingly in the low energy regime after several breaks of symmetries and before the 
electroweak break, the spin ($SO(1,3)$) and all 
the charges (the colour $SU(3)$, the weak $SU(2)$, the hyper charge $U(1)$) of quarks and leptons, 
left and right handed as assumed by the {\it standard model}.
$\,\,$ 
{\bf ii.} The second kind of the spin~\cite{snmb:hn02hn03},  
described by $\tilde{\gamma}^a$'s ($\{\tilde{\gamma}^a, \tilde{\gamma}^b\}_{+}= 2 \, \eta^{ab}$) and  
anticommuting with the Dirac $\gamma^a$ ($\{\gamma^a, \tilde{\gamma}^b\}_{+}=0$),  
defines the families of spinors, which at the symmetries of 
$\;SO(1,3) \times SU(2)_{I} \times SU(2)_{II} \times U(1)_{II} \times SU(3)$ manifest two %decoupled 
groups of four massless families.

There is no  third kind of the Clifford algebra objects. The appearance of the two kinds of the 
Clifford algebra objects can be understood as follows:
If the Dirac one corresponds to the multiplication of any spinor object $B$ (any product of the Dirac 
$\gamma^a$'s, which represents a spinor state when being applied on a spinor vacuum state $|\psi_0>$) 
from the left hand side, the second kind of the Clifford objects  can be  understood 
(up to a factor, determining the Clifford evenness ($n_B=2k$) or oddness ($n_B=2k+1$) of the object  
$B$ as the multiplication of the object from the right hand side
\begin{eqnarray}
 \tilde{\gamma}^a B \, |\psi_0>: = i(-)^{n_B} B \gamma^a\, |\psi_0>, 
 \label{Bt}
 \end{eqnarray}
 with $|\psi_0>$ determining the 
spinor vacuum state. 
Accordingly we have
\begin{eqnarray}
&& \{ \gamma^a, \gamma^b\}_{+} = 2\eta^{ab} =  
\{ \tilde{\gamma}^a, \tilde{\gamma}^b\}_{+},\quad
\{ \gamma^a, \tilde{\gamma}^b\}_{+} = 0,\nonumber\\
&&S^{ab}: = (i/4) (\gamma^a \gamma^b - \gamma^b \gamma^a), \quad
\tilde{S}^{ab}: = (i/4) (\tilde{\gamma}^a \tilde{\gamma}^b 
- \tilde{\gamma}^b \tilde{\gamma}^a),\quad  \{S^{ab}, \tilde{S}^{cd}\}_{-}=0.
\label{snmb:tildegclifford}
\end{eqnarray}
More detailed explanation can be found in  appendix~\ref{technique}. 
The  {\it spin-charge-family} theory proposes in $d=(1+13)$ a simple action for a Weyl 
spinor and for  the corresponding gauge fields  
\begin{eqnarray}
S            \,  &=& \int \; d^dx \; E\;{\mathcal L}_{f} +  
\nonumber\\  
               & & \int \; d^dx \; E\; (\alpha \,R + \tilde{\alpha} \, \tilde{R}),
               \end{eqnarray}
\begin{eqnarray}
{\mathcal L}_f &=& \frac{1}{2}\, (E\bar{\psi} \, \gamma^a p_{0a} \psi) + h.c., 
\nonumber\\
p_{0a }        &=& f^{\alpha}{}_a p_{0\alpha} + \frac{1}{2E}\, \{ p_{\alpha}, E f^{\alpha}{}_a\}_-, 
\nonumber\\  
   p_{0\alpha} &=&  p_{\alpha}  - 
                    \frac{1}{2}  S^{ab} \omega_{ab \alpha} - 
                    \frac{1}{2}  \tilde{S}^{ab}   \tilde{\omega}_{ab \alpha},                   
\nonumber\\ 
R              &=&  \frac{1}{2} \, \{ f^{\alpha [ a} f^{\beta b ]} \;(\omega_{a b \alpha, \beta} 
- \omega_{c a \alpha}\,\omega^{c}{}_{b \beta}) \} + h.c. \;, 
\nonumber\\
\tilde{R}      &=& \frac{1}{2}\,   f^{\alpha [ a} f^{\beta b ]} \;(\tilde{\omega}_{a b \alpha,\beta} - 
\tilde{\omega}_{c a \alpha} \tilde{\omega}^{c}{}_{b \beta}) + h.c.\;. 
\label{wholeaction}
\end{eqnarray}
Here~\footnote{$f^{\alpha}{}_{a}$ are inverted 
vielbeins to 
$e^{a}{}_{\alpha}$ with the properties $e^a{}_{\alpha} f^{\alpha}{\!}_b = \delta^a{\!}_b,\; 
e^a{\!}_{\alpha} f^{\beta}{\!}_a = \delta^{\beta}_{\alpha} $. 
Latin indices  
$a,b,..,m,n,..,s,t,..$ denote a tangent space (a flat index),
while Greek indices $\alpha, \beta,..,\mu, \nu,.. \sigma,\tau ..$ denote an Einstein 
index (a curved index). Letters  from the beginning of both the alphabets
indicate a general index ($a,b,c,..$   and $\alpha, \beta, \gamma,.. $ ), 
from the middle of both the alphabets   
the observed dimensions $0,1,2,3$ ($m,n,..$ and $\mu,\nu,..$), indices from 
the bottom of the alphabets
indicate the compactified dimensions ($s,t,..$ and $\sigma,\tau,..$). 
We assume the signature $\eta^{ab} =
diag\{1,-1,-1,\cdots,-1\}$.} 
$f^{\alpha [a} f^{\beta b]}= f^{\alpha a} f^{\beta b} - f^{\alpha b} f^{\beta a}$. 
To see that the action~(Eq.(\ref{wholeaction})) manifests  after the break of 
symmetries~\cite{pikanorma,Portoroz03,gmdn}
all the known gauge fields and the scalar fields and the mass matrices of the observed families,
let us rewrite formally the action for a Weyl spinor of~(Eq.(\ref{wholeaction})) % in $d=(1+13)$  
as follows  
\begin{eqnarray}
{\mathcal L}_f &=&  \bar{\psi}\gamma^{n} (p_{n}- \sum_{A,i}\; g^{A}\tau^{Ai} A^{Ai}_{n}) \psi 
+ \nonumber\\
               & &  \{ \sum_{s=7,8}\;  \bar{\psi} \gamma^{s} p_{0s} \; \psi \}  + \nonumber\\
               & & {\rm the \;rest}, 
\label{faction}
\end{eqnarray}
where $n=0,1,2,3$ with
\begin{eqnarray}
\tau^{Ai} = \sum_{a,b} \;c^{Ai}{ }_{ab} \; S^{ab},
\nonumber\\ 
\{\tau^{Ai}, \tau^{Bj}\}_- = i \delta^{AB} f^{Aijk} \tau^{Ak}.
\label{tau}
\end{eqnarray}
All the charge ($\tau^{Ai}$ (Eqs.~(\ref{tau}), (\ref{so4}), (\ref{so6})) 
and the spin (Eq.~(\ref{so13})) operators are expressible with $S^{ab}$, which determine all the 
internal degrees of freedom of one family.   
 
Index $A$ enumerates all possible spinor charges and $g^A$ is the coupling constant to a particular 
gauge vector field $A^{Ai}_{n}$.  
Before the  break from $SO(1,3) \times SU(2)_{I} \times SU(2)_{II} \times U(1)_{II} \times  SU(3)$ to
$SO(1,3) \times SU(2)_{I} \times U(1)_{I} \times SU(3)$, $\tau^{3i}$  describe the colour 
charge ($SU(3)$), $\tau^{1i}$ the weak charge ($SU(2)_{I}$),  
$\tau^{2i}$ the second $SU(2)_{II}$ charge and $\tau^{4}$ determines the $U(1)_{II}$ charge.  
 After the  break  of $SU(2)_{II} \times U(1)_{II}$ to $ U(1)_{I}$ stays 
$A=2$  for the  $U(1)_{I}$ hyper charge $Y$ and 
after the second break  of $SU(2)_{I} \times U(1)_{I}$ to $U(1)$ stays
$A=2$  for the electromagnetic  charge $Q$, while instead of the weak 
charge $Q'$ and $\tau^{\pm}$ of the {\it standard model} manifest.

The breaks of the starting symmetry from $SO(1,13)$ to the symmetry $SO(1,7) \times SU(3) \times U(1)_{II}$ 
and further to $SO(1,3) \times SU(2)_{I} \times SU(2)_{II} \times U(1)_{II} \times SU(3) $ 
are assumed to leave  the low lying eight families of spinors 
massless~\footnote{We proved that it is possible to 
have massless fermions after a (particular) break if we start with massless fermions and 
assume particular boundary conditions after the break or  the "effective 
two dimensionality" cases~\cite{snmb:hn02hn03,dhn,DHN}}. After the break of $SO(1,13)$ to 
$SO(1,7) \times SU(3) \times U(1)$ there are eight such families ($2^{8/2-1}$), all left 
handed with respect to $SO(1,13)$. 

Accordingly the first row of the action in Eq.~(\ref{faction}) manifests the  
dynamical fermion part of the action, while the second part manifests, 
when particular superposition of either $\omega_{ab \sigma}$  or of $\tilde{\omega}_{ab \sigma}$ 
($\sigma \in (7,8)$) or both 
fields gain  nonzero vacuum expectation values, the mass matrices of fermions on the tree level. 
Scalar fields contribute also to masses of those gauge fields, which at a particular break lose symmetries.  
It is assumed that the symmetries in the $\tilde{S}^{ab} \tilde{\omega}_{abc}$ and in the 
$S^{ab}\omega_{abc}$ part break in a correlated way, triggered by particular superposition of scalar 
(with respect to the rest of symmetry) vielbeins and spin connections of both kinds  ($\omega_{abc}$ 
and $\tilde{\omega}_{abc}$).
I comment this part in sections~\ref{yukawandhiggs},~\ref{yukawatreefbelow}.  
The Majorana term, manifesting the Majorana neutrinos, is contained in the second row as 
well~(\ref{Majoranas}).
The third row in Eq.~(\ref{faction}) stays for all the rest, which is expected to be at low energies  
negligible or might slightly influence the mass matrices beyond the tree level.

 The generators $\tilde{S}^{ab}$ (Eqs.~(\ref{so13}), (\ref{so4}), (\ref{so6})) transform each 
 member of one family into the corresponding member of another family, due to the fact that 
 $\{S^{ab}, \tilde{S}^{cd}\}_{-}=0$ (Eq.(\ref{snmb:tildegclifford},\ref{tildesabfam})). 
 
 Correspondingly the action for the vielbeins and the spin connections of $S^{ab}$, with 
 the Lagrange density $\alpha\,E\, R$,  manifests at  the low energy regime, after  
 breaks of the starting symmetry, as the known vector gauge fields -- the gauge fields of 
 $U(1), \,SU(2),\, SU(3)$ and 
 the ordinary gravity, while $\tilde{\alpha}\,E \, \tilde{R}$ are responsible for masses of the gauge fields 
and the mass matrices of fermions. 
 
 After the electroweak break the effective Lagrange density for spinors (is expected to) looks like
 \begin{eqnarray}
 {\mathcal L}_f &=&  \bar{\psi}\, (\gamma^{m} \, p_{0m} - \,M) \psi\, , \nonumber\\
          p_{0m}&=& p_{m} - \{ e \,Q\,A_{m} +  g^{1} \cos \theta_1 \,Q'\, Z^{Q'}_{m} +
          \frac{g^{1}}{\sqrt{2}}\, (\tau^{1+} \,W^{1+}_{m} + \tau^{1-} \,W^{1-}_{m}) + \nonumber\\
                &+& g^{2} \cos \theta_2 \,Y'\, A^{Y'}_{m} +
          \frac{g^{2}}{\sqrt{2}}\, (\tau^{2+} \,A^{2+}_{m} + \tau^{2-} \,A^{2-}_{m})\, , \nonumber\\
           \bar{\psi}\,  M \, \psi &=&  \bar{\psi}\, \gamma^{s} \, p_{0s} \psi\,\nonumber\\
          p_{0s}&=& p_{s} - \{ \tilde{g}^{\tilde{N}_R}\, \vec{\tilde{N}}_R \,\vec{\tilde{A}}^{\tilde{N}_R}_{s} +
 	                \tilde{g}^{\tilde{Y}'}  \, \tilde{Y}'\,\tilde{A}^{\tilde{Y}'}_{s}
 	       	          + \frac{\tilde{g}^{2}}{\sqrt{2}}\, (\tilde{\tau}^{2+} \,\tilde{A}^{2+}_{s}
 	          + \tilde{\tau}^{2-}\,  \tilde{A}^{2-}_{s})   \nonumber\\
 	       &+&  \tilde{g}^{\tilde{N}_L} \,\vec{\tilde{N}}_L\, \vec{\tilde{A}}^{\tilde{N}_L}_{s} +
 		  	                \tilde{g}^{\tilde{Q}'} \, \tilde{Q}'\,
 		  	                \tilde{A}^{\tilde{Q}'}_{s}
 		  	       	          + \frac{\tilde{g}^{1}}{\sqrt{2}}\, (\tilde{\tau}^{1+}
 		  	       	          \,\tilde{A}^{1+}_{s}
 	          + \tilde{\tau}^{1-}\,  \tilde{A}^{1-}_{s})\nonumber\\ 	
 	          &+&  e\, Q\, A_{s} +  g^{1}\, \cos \theta_1 \,Q'\, Z^{Q'}_{s} +
          %\frac{g^{1}}{\sqrt{2}}\, (\tau^{1+} W^{1+}_{s} + \tau^{1-} W^{1-}_{s}) +
               g^{2} \cos \theta_2\, Y'\, A^{Y'}_{s}\,\}\,,
 	          \; s\in\{7,8\}\,.
 \label{factionI}
 \end{eqnarray}
 The term $\bar{\psi}\,  M \, \psi $ determines the tree level mass matrices of quarks and leptons. 
 The contributions to the mass matrices appear  at two very different energy scales due to two separate breaks. 
 Before the break  of $SU(2)_{II} \times U(1)_{II}$ to  $ U(1)_{I}$  the vacuum expectation values 
 of the scalar fields  appearing in $p_{0s}$ are all zero. The corresponding 
 dynamical scalar fields are massless. All the eight families are  massless and  
 the vector gauge fields $A^{Ai}_{m}\,,\, A=2\,,1\,,4;$  in Eq.~(\ref{faction}) are massless as well. 
 To the  break of $SU(2)_{II} \times U(1)_{II}$ to  $ U(1)_{I}$  
 the scalar fields from the first row in the covariant momentum $p_{0s}$, that is  the two triplets 
 $\vec{\tilde{A}}^{\tilde{N}_R}_{s} $  and $\vec{\tilde{A}}^{2}_{s} $ are assumed to contribute, 
 gaining non zero vacuum expectation values. 
 The upper four families, which are doublets with respect to the  infinitesimal generators 
 of the corresponding groups, namely $\vec{\tilde{N}}_R$ and 
 $\vec{\tilde{\tau}}^{2}$, become massive. No scalar fields of the kind  $
\omega_{abs}$ is assumed to contribute in this break. Therefore,  
 the lower four families, which are singlets with 
 respect to $\vec{\tilde{N}}_R$ and  $\vec{\tilde{\tau}}^{2}$, stay massless. 
 Due to the break of $SU(2)_{II} \times U(1)_{II}$ symmetries in the space of $\tilde{S}^{ab}$ and 
 $S^{ab}$,  the gauge fields $\vec{A}^{2}_{m}$ become massive. 
 The gauge vector fields $\vec{A}^{1}_{m}$  and $\vec{A}^{Y}_{m}$ stay massless at this break.

  To the  break of $SU(2)_{I} \times U(1)_{I}$ to  $ U(1)$  
  the scalar fields from the second row in the covariant momentum $p_{0s}$, that is the triplets 
  $\vec{\tilde{A}}^{\tilde{N}_L}_{s} $ and   $\vec{\tilde{A}}^{1}_{s} $ and the singlet 
  $\tilde{A}^{4}_{s}$,  as well as the ones 
  from the third row originating in $\omega_{abc}$, that is ($A_{s}$, $ Z^{Q'}_{s}$, $ A^{Y'}_{s}$), 
  are assumed to contribute, 
 by gaining non zero vacuum expectation values. 
 
 This electroweak break causes  non zero mass matrices of the lower four families. 
 Also the gauge fields $Z^{Q'}_{m}$, $ W^{1+}_{m}$ and $ W^{1-}_{m}$ gain masses. 
  The electroweak break influences slightly the mass matrices %on the tree level  
 of the   upper four families, due to 
  the contribution of $A_{s}$, $ Z^{Q'}_{s}$, $ A^{Y'}_{s}$ and $\tilde{A}^{4}_{s}$
  and in loop corrections also $ Z^{Q'}_{m}$ and $W^{\pm}_{m}$.

 To loops corrections of both groups of families the massive vector gauge fields contribute. 
 The dynamical massive scalar fields contribute only to families of the group to 
 which the couple.

 The detailed explanation of the two phase transitions which manifest in 
 Eq.~(\ref{factionI}) is presented in what follows.

\subsection{Spinor action through  breaks}
\label{spinoraction}

In this subsection properties of quarks, $u$ and $d$, and leptons, $\nu$ and $e$, of two 
groups of four  families are presented, before and after the two successive 
breaks, in which eight massless families gain their masses. At the break of $SO(1,3) \times SU(2)_{I} 
\times SU(2)_{II} \times U(1)_{II} \times SU(3)$ to $SO(1,3) \times SU(2)_{I} \times U(1)_{I} 
\times SU(3)$ the four families coupled to the scalar fields which gain at this break nonzero 
vacuum expectation values become massive, while the  four families which do not couple to 
these scalar fields stay massless, representing four families of left handed weak charged 
colour triplets quarks ($u_{L}$, $d_{L}$), right handed weak chargeless colour triplets 
quarks ($u_{R}$, $d_{R}$), left handed weak charged colour singlets leptons ($\nu_{L}$, $e_{L}$) 
and right handed weak chargeless colour singlets leptons ($\nu_{R}$, $e_{R}$). The lowest 
of the upper four families, after the second break having all the properties of the 
{\it standard model} quarks and leptons except the masses, are candidates to form the 
dark matter clusters.

After the second break from $SO(1,3) \times SU(2)_{I} \times U(1)_{I} \times SU(3)$ to 
$SO(1,3) \times \times U(1) \times SU(3)$ the  last four families become massive due 
to the nonzero vacuum expectation values of the rest of scalar fields. Three of them, 
represent the observed families of quarks and leptons, so far included into the 
{\it standard model}, if we do not count the right handed $\nu$. 

The technique~\cite{snmb:hn02hn03}, which offers an easy way to keep a track of 
the symmetry properties of spinors, is used as a tool to clearly demonstrate properties of 
spinors. This technique is  explained in more details in 
appendix~\ref{technique}. In this subsection only a short introduction, needed to 
follow the explanation, is presented. Mass matrices of each groups of four families, on 
the tree and below the tree level, originated in the scalar gauge fields, which at each of 
the two breaks gain a nonzero vacuum expectation values, will be discussed in  
section~\ref{yukawatreefbelow}. 

Following the refs.~\cite{snmb:hn02hn03} we define nilpotents ($\,\stackrel{ab}{(k)}^2= 0$) 
and projectors ($\,\stackrel{ab}{[k]}^2 = \stackrel{ab}{[k]}$) (Eq.~(\ref{signature}) in 
appendix~\ref{technique})   
\begin{eqnarray}
\stackrel{ab}{(\pm i)}: &=& \frac{1}{2}(\gamma^a \mp  \gamma^b),  \; 
\stackrel{ab}{[\pm i]}: = \frac{1}{2}(1 \pm \gamma^a \gamma^b), \quad
{\rm for} \,\; \eta^{aa} \eta^{bb} = -1, \nonumber\\
\stackrel{ab}{(\pm )}: &= &\frac{1}{2}(\gamma^a \pm i \gamma^b),  \; 
\stackrel{ab}{[\pm ]}: = \frac{1}{2}(1 \pm i\gamma^a \gamma^b), \quad
{\rm for} \,\; \eta^{aa} \eta^{bb} =1\,,
\label{snmb:eigensab}
\end{eqnarray} 
as  eigenvectors  of $S^{ab}$ as well as of $\tilde{S}^{ab}$ (Eq.~(\ref{grapheigen}) in 
appendix~\ref{technique})   
\begin{eqnarray}
S^{ab} \stackrel{ab}{(k)} =  \frac{k}{2} \stackrel{ab}{(k)}, \quad 
S^{ab} \stackrel{ab}{[k]} =  \frac{k}{2} \stackrel{ab}{[k]}, \quad
\tilde{S}^{ab} \stackrel{ab}{(k)}  = \frac{k}{2} \stackrel{ab}{(k)},  \quad 
\tilde{S}^{ab} \stackrel{ab}{[k]}  =   - \frac{k}{2} \stackrel{ab}{[k]}.\;\;
\label{snmb:eigensabev}
\end{eqnarray}
One can easily verify that $\gamma^a$ transform $\stackrel{ab}{(k)}$ into  $\stackrel{ab}{[-k]}$, 
while $\tilde{\gamma}^a$ transform  $\stackrel{ab}{(k)}$ into $\stackrel{ab}{[k]}$ 
(Eq.~(\ref{snmb:gammatildegamma}) in appendix~\ref{technique})   
\begin{eqnarray}
\gamma^a \stackrel{ab}{(k)}= \eta^{aa}\stackrel{ab}{[-k]},\; 
\gamma^b \stackrel{ab}{(k)}= -ik \stackrel{ab}{[-k]}, \; 
\gamma^a \stackrel{ab}{[k]}= \stackrel{ab}{(-k)},\; 
\gamma^b \stackrel{ab}{[k]}= -ik \eta^{aa} \stackrel{ab}{(-k)},\;\;\;
\label{snmb:graphgammaaction}
\end{eqnarray}
\begin{eqnarray}
\tilde{\gamma^a} \stackrel{ab}{(k)} = - i\eta^{aa}\stackrel{ab}{[k]},\;
\tilde{\gamma^b} \stackrel{ab}{(k)} =  - k \stackrel{ab}{[k]}, \;
\tilde{\gamma^a} \stackrel{ab}{[k]} =  \;\;i\stackrel{ab}{(k)},\; 
\tilde{\gamma^b} \stackrel{ab}{[k]} =  -k \eta^{aa} \stackrel{ab}{(k)}. \; 
\label{snmb:gammatilde}
\end{eqnarray}
Correspondingly, $\tilde{S}^{ab}$ generate the equivalent representations to 
representations of $S^{ab}$,  and opposite.
Defining  the basis vectors in the internal space of 
spin degrees of freedom 
in $d=(1+13)$  as products of  projectors and nilpotents from Eq.~(\ref{snmb:eigensab}) 
on the spinor vacuum state $|\psi_0>$, the representation of one Weyl spinor with respect to 
$S^{ab}$  manifests after the breaks  
the spin and all the charges of one family members, and the gauge fields of $\tilde{S}^{ab}$ 
manifest as all the observed gauge fields. $\tilde{S}^{ab}$ determine families and 
correspondingly the family quantum numbers, while scalar gauge fields of $\tilde{S}^{ab}$ determine,  
together with particular scalar gauge fields of $S^{ab}$, mass matrices, manifesting 
effectively as Yukawa fields and Higgs. 

Expressing the operators $\gamma^7$ and $\gamma^8$ in terms of the nilpotents $\stackrel{78}{(\pm)}$, the
mass term in Eqs.~(\ref{faction}, \ref{factionI}) can be rewritten as follows
\begin{eqnarray}
\bar{\psi} M \psi   &=& \sum_{s=7,8}\,\bar{\psi} \gamma^{s}\, p_{0s}\,\psi  =
 \psi^{\dagger}\, \gamma^{0}\, (\stackrel{78}{(-)}\,p_{0-} + \stackrel{78}{(+)}\,p_{0+}) \psi\, ,\nonumber\\ 
\stackrel{78}{(\pm)}&=&  \frac{1}{2}\,(\gamma^{7} \, \pm i\,\gamma^{8} )\,, \nonumber\\ 
p_{0\pm}&=& (p_{07} \mp i\, p_{08})\,. 
\label{factionM}
\end{eqnarray}

After the breaks  of the starting symmetry   
(from  $SO(1,13)$ through  $SO(1,7) \times U(1)_{II} \times SU(3)$) to $SO(1,3) \times 
SU(2)_{I} \times SU(2)_{II} \times U(1)_{II} \times SU(3)$ there are  eight ($2^{\frac{8}{2}-1}$) massless 
families of spinors. (Some support for this assumption is made when studying 
toy models~\cite{dhn,DHN}.) 

Family members  $\alpha \in (u\,,d\,,\nu\,,e)$ carry the $U(1)_{II}\,$  charge (the generator 
of the infinitesimal transformations of the group is $\tau^4$, 
presented in Eq.~(\ref{so6})), the $ \,SU(3)$ charge (the generators are $\vec{\tau}^{3}$, presented in 
Eq.~(\ref{so6}))  and the two $SU(2)$ charges,  $SU(2)_{II}$ and $SU(2)_{I}$ (the generators are 
presented in Eq.~(\ref{so4}) as $\vec{\tau}^2$  and $\vec{\tau}^1$, respectively). Family members 
are in the representations, in which the left handed (with respect to  $SO(1,3)$)  
carry the $SU(2)_{I}$ (weak) charge (with the corresponding generators $\vec{\tau}^1$), while   
the  right handed  carry the $SU(2)_{II}$  charge (with the corresponding generators $\vec{\tau}^2$).

Each family member carries also the  family quantum number, which concern  $\tilde{S}^{ab}$ and is 
determined by the quantum numbers of the two $SU(2)$ from $SO(1,3)$ (with the generators 
$\vec{\tilde{N}}_{(L,R)}$, Eq.~(\ref{so6})) and the two $SU(2)$ from $SO(4)$ (with the generators 
$\vec{\tilde{\tau}}^{(1,2)}$, Eq.~(\ref{so4})). 

Properties of families of spinors can transparently be analysed if using our technique. 
We arrange products of nilpotents and projectors to be  eigenvectors of 
the Cartan subalgebra $S^{03}, S^{12}, S^{56}, S^{78}, S^{9\,10}, S^{11\,12}, S^{13\,14}$ and, 
at the same time,  they are also the eigenvectors of the corresponding  $\tilde{S}^{ab}$, 
that is of $\tilde{S}^{03}, \tilde{S}^{12}, \tilde{S}^{56}, \tilde{S}^{78}, \tilde{S}^{9\,10}, 
\tilde{S}^{11\,12}, \tilde{S}^{13\,14}$.

Below the generators of the infinitesimal transformations of the subgroups of 
the group  $SO(1,13)$ in the $S^{ab}$ and $\tilde{S}^{ab}$ sectors,  responsible for the 
properties of spinors in the low energy regime, are presented.  
\begin{eqnarray}
\label{so13}
\vec{N}_{\pm}(= \vec{N}_{(L,R)}): &=& \,\frac{1}{2} (S^{23}\pm i S^{01},S^{31}\pm i S^{02}, 
S^{12}\pm i S^{03} )\,\,,\nonumber\\
\vec{\tilde{N}}_{\pm}(=\vec{\tilde{N}}_{(L,R)}):&=& \,\frac{1}{2} (\tilde{S}^{23}\pm i \tilde{S}^{01},
\tilde{S}^{31}\pm i \tilde{S}^{02}, \tilde{S}^{12}\pm i \tilde{S}^{03} )\,
\end{eqnarray}
determine representations of the two $SU(2)$ subgroups of $SO(1,3)$, 
 \begin{eqnarray}
 \label{so4}
 \vec{\tau}^{1}:&=&\frac{1}{2} (S^{58}-  S^{67}, \,S^{57} + S^{68}, \,S^{56}-  S^{78} )\,,\;\;
 \vec{\tau}^{2}:= \frac{1}{2} (S^{58}+  S^{67}, \,S^{57} - S^{68}, \,S^{56}+  S^{78} )\,,\,\;\nonumber\\
 \vec{\tilde{\tau}}^{1}:&=&\frac{1}{2} (\tilde{S}^{58}-  \tilde{S}^{67}, \,\tilde{S}^{57} + 
 \tilde{S}^{68}, \,\tilde{S}^{56}-  \tilde{S}^{78} )\,,\;\;
 \vec{\tilde{\tau}}^{2}:=\frac{1}{2} (\tilde{S}^{58}+  \tilde{S}^{67}, \,\tilde{S}^{57} - 
 \tilde{S}^{68}, \,\tilde{S}^{56}+  \tilde{S}^{78} ),\,\,\;\;
 \end{eqnarray}
 determine representations of $SU(2)_{I}\times$ $SU(2)_{II}$ of $SO(4)$
 and 
  \begin{eqnarray}
 \label{so6}
 \vec{\tau}^{3}: = &&\frac{1}{2} \,\{  S^{9\;12} - S^{10\;11} \,,
  S^{9\;11} + S^{10\;12} ,\, S^{9\;10} - S^{11\;12} ,\nonumber\\
 && S^{9\;14} -  S^{10\;13} ,\,  S^{9\;13} + S^{10\;14} \,,
  S^{11\;14} -  S^{12\;13}\,,\nonumber\\
 && S^{11\;13} +  S^{12\;14} ,\, 
 \frac{1}{\sqrt{3}} ( S^{9\;10} + S^{11\;12} - 
 2 S^{13\;14})\}\,,\nonumber\\
 \tau^{4}: = &&-\frac{1}{3}(S^{9\;10} + S^{11\;12} + S^{13\;14})\,,\;\;
 \tilde{\tau}^{4}: = -\frac{1}{3}(\tilde{S}^{9\;10} + \tilde{S}^{11\;12} + \tilde{S}^{13\;14})\,,
 \end{eqnarray}
 determine representations of $SU(3) \times U(1)$,  originating in $SO(6)$. 
 
 It is assumed that  at the break of $SO(1,13)$ to $SO(1,7)\times U(1)_{II}\times SU(3)$ 
 all spinors but one become massive, which then manifests eight massless families 
 generated by those generators of the infinitesimal transformations $\tilde{S}^{ab}$ 
 which belong to the subgroup $SO(1,7)$. Some justification for such an assumption 
 can be found in the refs.~\cite{dhn,DHN}.

At the stage of the symmetry $\;SO(1,3) \times SU(2)_{I} \times SU(2)_{II} \times U(1)_{II} \times SU(3)$ 
each member of a family appears in eight massless families. Each family manifests at this symmetry 
eightplets  of $u$ and $d$ quarks, left handed weak charged and right handed weak chargeless (of spin 
 ($\pm\frac{1}{2}$)), in three colours and the colourless eightplet of $\nu$ and $e$ leptons, 
 left handed weak charged and right handed weak chargeless (of spin ($\pm\frac{1}{2}$)). 

In Table~\ref{Table I.} the eightplet of quarks of a particular colour charge ($\tau^{33}=1/2$, 
$\tau^{38}=1/(2\sqrt{3})$) and the $U(1)_{II}$ charge ($\tau^{4}=1/6$) is presented in our 
technique~\cite{snmb:hn02hn03}, as products of nilpotents and projectors. 
\begin{table}
\begin{center}
\begin{tabular}{|r|c||c||c|c||c|c|c||r|r|}
\hline
i&$$&$|^a\psi_i>$&$\Gamma^{(1,3)}$&$ S^{12}$&$\Gamma^{(4)}$&
$\tau^{13}$&$\tau^{23}$&$Y$&$Q$\\
\hline\hline
&& ${\rm Octet},\;\Gamma^{(1,7)} =1,\;\Gamma^{(6)} = -1,$&&&&&&& \\
&& ${\rm of \; quarks}$&&&&&&&\\
\hline\hline
1&$ u_{R}^{c1}$&$ \stackrel{03}{(+i)}\,\stackrel{12}{(+)}|
\stackrel{56}{(+)}\,\stackrel{78}{(+)}
||\stackrel{9 \;10}{(+)}\;\;\stackrel{11\;12}{[-]}\;\;\stackrel{13\;14}{[-]} $
&1&$\frac{1}{2}$&1&0&$\frac{1}{2}$&$\frac{2}{3}$&$\frac{2}{3}$\\
\hline 
2&$u_{R}^{c1}$&$\stackrel{03}{[-i]}\,\stackrel{12}{[-]}|\stackrel{56}{(+)}\,\stackrel{78}{(+)}
||\stackrel{9 \;10}{(+)}\;\;\stackrel{11\;12}{[-]}\;\;\stackrel{13\;14}{[-]}$
&1&$-\frac{1}{2}$&1&0&$\frac{1}{2}$&$\frac{2}{3}$&$\frac{2}{3}$\\
\hline
3&$d_{R}^{c1}$&$\stackrel{03}{(+i)}\,\stackrel{12}{(+)}|\stackrel{56}{[-]}\,\stackrel{78}{[-]}
||\stackrel{9 \;10}{(+)}\;\;\stackrel{11\;12}{[-]}\;\;\stackrel{13\;14}{[-]}$
&1&$\frac{1}{2}$&1&0&$-\frac{1}{2}$&$-\frac{1}{3}$&$-\frac{1}{3}$\\
\hline 
4&$ d_{R}^{c1} $&$\stackrel{03}{[-i]}\,\stackrel{12}{[-]}|
\stackrel{56}{[-]}\,\stackrel{78}{[-]}
||\stackrel{9 \;10}{(+)}\;\;\stackrel{11\;12}{[-]}\;\;\stackrel{13\;14}{[-]} $
&1&$-\frac{1}{2}$&1&0&$-\frac{1}{2}$&$-\frac{1}{3}$&$-\frac{1}{3}$\\
\hline
5&$d_{L}^{c1}$&$\stackrel{03}{[-i]}\,\stackrel{12}{(+)}|\stackrel{56}{[-]}\,\stackrel{78}{(+)}
||\stackrel{9 \;10}{(+)}\;\;\stackrel{11\;12}{[-]}\;\;\stackrel{13\;14}{[-]}$
&-1&$\frac{1}{2}$&-1&$-\frac{1}{2}$&0&$\frac{1}{6}$&$-\frac{1}{3}$\\
\hline
6&$d_{L}^{c1} $&$\stackrel{03}{(+i)}\,\stackrel{12}{[-]}|
\stackrel{56}{[-]}\,\stackrel{78}{(+)}
||\stackrel{9 \;10}{(+)}\;\;\stackrel{11\;12}{[-]}\;\;\stackrel{13\;14}{[-]} $
&-1&$-\frac{1}{2}$&-1&$-\frac{1}{2}$&0&$\frac{1}{6}$&$-\frac{1}{3}$\\
\hline
7&$ u_{L}^{c1}$&$\stackrel{03}{[-i]}\,\stackrel{12}{(+)}|
\stackrel{56}{(+)}\,\stackrel{78}{[-]}
||\stackrel{9 \;10}{(+)}\;\;\stackrel{11\;12}{[-]}\;\;\stackrel{13\;14}{[-]}$
&-1&$\frac{1}{2}$&-1&$\frac{1}{2}$&0&$\frac{1}{6}$&$\frac{2}{3}$\\
\hline
8&$u_{L}^{c1}$&$\stackrel{03}{(+i)}\,\stackrel{12}{[-]}|\stackrel{56}{(+)}\,\stackrel{78}{[-]}
||\stackrel{9 \;10}{(+)}\;\;\stackrel{11\;12}{[-]}\;\;\stackrel{13\;14}{[-]}$
&-1&$-\frac{1}{2}$&-1&$\frac{1}{2}$&0&$\frac{1}{6}$&$\frac{2}{3}$\\
\hline\hline
\end{tabular}
\end{center}
\caption{\label{Table I.} The 8-plet of quarks - the members of $SO(1,7)$ subgroup of the 
group $SO(1,13)$, belonging to one Weyl left 
handed ($\Gamma^{(1,13)} = -1 = \Gamma^{(1,7)} \times \Gamma^{(6)}$) spinor representation of 
$SO(1,13)$ is presented in the technique~\cite{snmb:hn02hn03}. 
It contains the left handed weak charged quarks and the right handed weak chargeless quarks 
of a particular 
colour $(1/2,1/(2\sqrt{3}))$. Here  $\Gamma^{(1,3)}$ defines the handedness in $(1+3)$ space, 
$ S^{12}$ defines the ordinary spin (which can also be read directly from the basic vector, both
vectors  with both spins, $\pm \frac{1}{2}$, are presented), 
$\tau^{13}$ defines the third component of the weak charge, $\tau^{23}$ the third component 
of the $SU(2)_{II}$ charge, 
$\tau^{4}$ (the $U(1)$ charge) defines together with 
$\tau^{23}$  the hyper charge ($Y= \tau^4 + \tau^{23}$), $Q= Y + \tau^{13}$ is the 
electromagnetic charge. The vacuum state $|\psi_0>$, on which the nilpotents and 
projectors operate, is not shown. The basis is the massless one. 
The reader can find the whole Weyl representation in the ref.~\cite{Portoroz03}.}
\end{table}

In Table~\ref{Table II.} the eightplet of the colourless leptons of the $U(1)_{II}$ 
charge ($\tau^{4}=-1/2$) is presented in the same technique. 
\begin{table}
\begin{center}
\begin{tabular}{|r|c||c||c|c||c|c|c||r|r|}
\hline
i&$$&$|^a\psi_i>$&$\Gamma^{(1,3)}$&$ S^{12}$&$\Gamma^{(4)}$&
$\tau^{13}$&$\tau^{23}$&$Y$&$Q$\\
\hline\hline
&& ${\rm Octet},\;\Gamma^{(1,7)} =1,\;\Gamma^{(6)} = -1,$&&&&&&& \\
&& ${\rm of \; quarks}$&&&&&&&\\
\hline\hline
1&$ \nu_{R}$&$ \stackrel{03}{(+i)}\,\stackrel{12}{(+)}|
\stackrel{56}{(+)}\,\stackrel{78}{(+)}
||\stackrel{9 \;10}{(+)}\;\;\stackrel{11\;12}{(+)}\;\;\stackrel{13\;14}{(+)} $
&1&$\frac{1}{2}$&1&0&$\frac{1}{2}$&$0$&$0$\\
\hline 
2&$\nu_{R}$&$\stackrel{03}{[-i]}\,\stackrel{12}{[-]}|\stackrel{56}{(+)}\,\stackrel{78}{(+)}
||\stackrel{9 \;10}{(+)}\;\;\stackrel{11\;12}{(+)}\;\;\stackrel{13\;14}{(+)} $
&1&$-\frac{1}{2}$&1&0&$\frac{1}{2}$&$0$&$0$\\
\hline
3&$e_{R}$&$\stackrel{03}{(+i)}\,\stackrel{12}{(+)}|\stackrel{56}{[-]}\,\stackrel{78}{[-]}
||\stackrel{9 \;10}{(+)}\;\;\stackrel{11\;12}{(+)}\;\;\stackrel{13\;14}{(+)} $
&1&$\frac{1}{2}$&1&0&$-\frac{1}{2}$&$-1$&$-1$\\
\hline 
4&$ e_{R} $&$\stackrel{03}{[-i]}\,\stackrel{12}{[-]}|
\stackrel{56}{[-]}\,\stackrel{78}{[-]}
||\stackrel{9 \;10}{(+)}\;\;\stackrel{11\;12}{(+)}\;\;\stackrel{13\;14}{(+)} $
&1&$-\frac{1}{2}$&1&0&$-\frac{1}{2}$&$-1$&$-1$\\
\hline
5&$e_{L}$&$\stackrel{03}{[-i]}\,\stackrel{12}{(+)}|\stackrel{56}{[-]}\,\stackrel{78}{(+)}
||\stackrel{9 \;10}{(+)}\;\;\stackrel{11\;12}{(+)}\;\;\stackrel{13\;14}{(+)} $
&-1&$\frac{1}{2}$&-1&$-\frac{1}{2}$&0&$-\frac{1}{2}$&$-1$\\
\hline
6&$e_{L} $&$\stackrel{03}{(+i)}\,\stackrel{12}{[-]}|
\stackrel{56}{[-]}\,\stackrel{78}{(+)}
||\stackrel{9 \;10}{(+)}\;\;\stackrel{11\;12}{(+)}\;\;\stackrel{13\;14}{(+)} $
&-1&$-\frac{1}{2}$&-1&$-\frac{1}{2}$&0&$-\frac{1}{2}$&$-1$\\
\hline
7&$ \nu_{L}$&$ \stackrel{03}{[-i]}\,\stackrel{12}{(+)}|
\stackrel{56}{(+)}\,\stackrel{78}{[-]}
||\stackrel{9 \;10}{(+)}\;\;\stackrel{11\;12}{(+)}\;\;\stackrel{13\;14}{(+)} $
&-1&$\frac{1}{2}$&-1&$\frac{1}{2}$&0&$-\frac{1}{2}$&$0$\\
\hline
8&$\nu_{L}$&$\stackrel{03}{(+i)}\,\stackrel{12}{[-]}|\stackrel{56}{(+)}\,\stackrel{78}{[-]}
||\stackrel{9 \;10}{(+)}\;\;\stackrel{11\;12}{(+)}\;\;\stackrel{13\;14}{(+)} $
&-1&$-\frac{1}{2}$&-1&$\frac{1}{2}$&0&$-\frac{1}{2}$&$0$\\
\hline\hline
\end{tabular}
\end{center}
\caption{\label{Table II.} The 8-plet of leptons - the members of $SO(1,7)$ subgroup of the 
group $SO(1,13)$, 
belonging to one Weyl left 
handed ($\Gamma^{(1,13)} = -1 = \Gamma^{(1,7)} \times \Gamma^{(6)}$) spinor representation of 
$SO(1,13)$ is presented in the massless basis. 
It contains the colour chargeless left handed weak charged leptons and the right handed weak 
chargeless leptons. The rest of notation is the same as in Table~\ref{Table I.}.  
 %Here  $\Gamma^{(1,3)}$ defines the handedness in $(1+3)$ space, 
%$ S^{12}$ defines the ordinary spin (which can also be read directly from the basic vector, both
%vectors  with both spins, -$\pm \frac{1}{2}$, are presented), 
%$\tau^{13}$ defines the third component of the weak charge, $\tau^{23}$ the third component 
%of the $SU(2)_{II}$ charge, $\tau^{33}$ and 
%$\tau^{38}$ define the colour charge and 
%$\tau^{4}$ the $U(1)$ charge, which together with the
%$\tau^{23}$ defines the hyper charge $Y= \tau^4 + \tau^{23}$. 
%The reader can find the whole Weyl representation in the ref.~\cite{Portoroz03}.
}
\end{table}

In both tables the vectors are chosen to be the eigenvectors of the operators of 
handedness $\Gamma^{(n)}$ and $\tilde{\Gamma}^{(n)}$,  
the generators $\tau^{13}$ (the member of the weak $SU(2)_{I}$), $\, \tau^{23},$ (the member of 
 $SU(2)_{II}$), $ \,\tau^{33}\,,$  $ \tau^{38}$ (the members of $SU(3)$),  $Y\,(= \tau^{4} + \tau^{23})$ and
$Q \, (= Y + \tau^{13})$. They are also 
eigenvectors of the corresponding $\tilde{S}^{ab}$, $\tilde{\tau}^{Ai}, A=1,2,4$ and 
$\tilde{Y}$ and $ \tilde{Q}$. 
The  tables  for the two additional choices of the colour charge of quarks 
 follow from Table~\ref{Table I.} by changing the colour part of 
 the states~\cite{Portoroz03}, that is by applying $\tau^{3i}$, which are not 
 members of the Cartan subalgebra, on the states of Table~\ref{Table I.}.   

Looking at Tables~(\ref{Table I.}, \ref{Table II.}) and taking into account the relation 
$\stackrel{78}{(-)} \stackrel{78}{(+)}= - \,  \stackrel{78}{[-]}$ from Eq.~(\ref{graphbinoms}) in 
appendix~{\ref{technique}} and the relation 
$\gamma^{0}\, \stackrel{03}{(+i)}=  \stackrel{03}{[-i]}$ from Eq.~(\ref{snmb:graphgammaaction})
one  notices that the operator 
$\gamma^{0}\, \stackrel{78}{(-)}\,$ (Eq.(\ref{factionM})) transforms the right handed 
$u^{c1}_{R}$ from the first row of Table~\ref{Table I.} into the left handed $u^{c1}_{L}$  of 
the same spin and charge 
from the seventh row of the same table, and that it transform the right handed 
$\nu_{R}$ from the first row  of  Table~\ref{Table II.} into left handed $\nu_{L}$ from 
the seventh row of the same table, just what the Higgs and $\gamma^0$ do in the {\it standard model}. 
Equivalently one finds that the operator 
$\gamma^{0}\, \stackrel{78}{(+)}\,$ 
transforms the right handed $d^{c1}_{R}$-quark from the third row into the left handed one (of the same 
spin and colour) from the fifth row of Table~\ref{Table I.} and that it 
transforms the right handed $e_{R}$ from the third row of Table~\ref{Table II.} into the left 
handed one (of the same 
spin) from the fifth row of Table~\ref{Table II.}.

 The superposition of generators $\tilde{S}^{ab}$ forming  eight generators ($\tilde{N}^{\pm}_{R,L}$, 
 $\tilde{\tau}^{(2,1)\pm}$)  presented in appendix~\ref{technique},  Eq.~(\ref{plusminus}), 
 generate families, transforming each member of one family  into the same member of another family, 
 due to the fact that  $\{S^{ab}, \tilde{S}^{cd}\}_{-}=0$ (Eq.(\ref{snmb:tildegclifford})). 
 The eight families of the first 
 member of the eightplet of quarks from Table~\ref{Table I.}, for example, that is of the right 
 handed $u^{c1}_{R}$-quark  with spin $\frac{1}{2}$,  are presented in the left column of 
 Table~\ref{Table III.}. 
 The  generators ($\tilde{N}^{\pm}_{R,L}$, $\tilde{\tau}^{(2,1)\pm}$)  (Eq.~(\ref{plusminus})) 
 transform the first member of the eightplet from Table~\ref{Table II.},
 that is the right handed neutrino $\nu_{R}$ with  spin $\frac{1}{2}$, into the eight-plet of  
 right handed neutrinos with spin up, belonging to eight different families. These families are presented  
 in the right column of the same table. All the other members of any of the eight families of quarks or 
 leptons follow  from any member of a particular family by the application of the 
 operators ($N^{\pm}_{R,L}$, $\tau^{(2,1)\pm}$) on this particular member.  

 \begin{table}
 \begin{center}
 \begin{tabular}{|r||c||c||c||c||}
 \hline
 $I_R$ & $u_{R}^{c1}$&
  $ \stackrel{03}{(+i)}\,\stackrel{12}{[+]}|\stackrel{56}{[+]}\,\stackrel{78}{(+)}||
  \stackrel{9 \;10}{(+)}\;\;\stackrel{11\;12}{[-]}\;\;\stackrel{13\;14}{[-]}$ & 
  $\nu_{R}$&
  $ \stackrel{03}{[+i]}\,\stackrel{12}{(+)}|\stackrel{56}{[+]}\,\stackrel{78}{(+)}|| 
  \stackrel{9 \;10}{(+)}\;\;\stackrel{11\;12}{(+)}\;\;\stackrel{13\;14}{(+)}$ 
 \\
 \hline
  $II_R$ & $u_{R}^{c1}$&
  $ \stackrel{03}{[+i]}\,\stackrel{12}{(+)}|\stackrel{56}{[+]}\,\stackrel{78}{(+)}||
  \stackrel{9 \;10}{(+)}\;\;\stackrel{11\;12}{[-]}\;\;\stackrel{13\;14}{[-]}$ & 
  $\nu_{R}$&
  $ \stackrel{03}{(+i)}\,\stackrel{12}{[+]}|\stackrel{56}{(+)}\,\stackrel{78}{[+]}||
  \stackrel{9 \;10}{(+)}\;\;\stackrel{11\;12}{(+)}\;\;\stackrel{13\;14}{(+)}$ 
 \\
 \hline
 $III_R$ & $u_{R}^{c1}$&
 $ \stackrel{03}{(+i)}\,\stackrel{12}{[+]}|\stackrel{56}{(+)}\,\stackrel{78}{[+]}||
 \stackrel{9 \;10}{(+)}\;\;\stackrel{11\;12}{[-]}\;\;\stackrel{13\;14}{[-]}$ & 
 $\nu_{R}$&
 $ \stackrel{03}{(+i)}\,\stackrel{12}{[+]}|\stackrel{56}{[+]}\,\stackrel{78}{(+)}||
 \stackrel{9 \;10}{(+)}\;\;\stackrel{11\;12}{(+)}\;\;\stackrel{13\;14}{(+)}$ 
 \\
 \hline
 $IV_R$ & $u_{R}^{c1}$&
  $ \stackrel{03}{[+i]}\,\stackrel{12}{(+)}|\stackrel{56}{(+)}\,\stackrel{78}{[+]}||
  \stackrel{9 \;10}{(+)}\:\; \stackrel{11\;12}{[-]}\;\;\stackrel{13\;14}{[-]}$ & 
  $\nu_{R}$&
  $ \stackrel{03}{[+i]}\,\stackrel{12}{(+)}|\stackrel{56}{(+)}\,\stackrel{78}{[+]}||
  \stackrel{9 \;10}{(+)}\;\;\stackrel{11\;12}{(+)}\;\;\stackrel{13\;14}{(+)}$ 
 \\
 \hline\hline\hline
 $V_R$ & $u_{R}^{c1}$&
 $ \stackrel{03}{(+i)}\,\stackrel{12}{(+)}|\stackrel{56}{(+)}\,\stackrel{78}{(+)} ||
 \stackrel{9 \;10}{(+)}\;\;\stackrel{11\;12}{[-]}\;\;\stackrel{13\;14}{[-]}$ & 
 $\nu_{R}$&
 $ \stackrel{03}{(+i)}\,\stackrel{12}{(+)}|\stackrel{56}{(+)}\,\stackrel{78}{(+)} ||
 \stackrel{9 \;10}{(+)}\;\;\stackrel{11\;12}{(+)}\;\;\stackrel{13\;14}{(+)}$ 
 \\
 \hline
 $V_R$ & $u_{R}^{c1}$&
  $ \stackrel{03}{[+i]}\,\stackrel{12}{[+]}|\stackrel{56}{[+]}\,\stackrel{78}{[+]}|| 
  \stackrel{9 \;10}{(+)}\;\;\stackrel{11\;12}{[-]}\;\;\stackrel{13\;14}{[-]}$ & 
  $\nu_{R}$&
  $ \stackrel{03}{[+i]}\,\stackrel{12}{[+]}|\stackrel{56}{[+]}\,\stackrel{78}{[+]}|| 
 \stackrel{9 \;10}{(+)}\;\;\stackrel{11\;12}{(+)}\;\;\stackrel{13\;14}{(+)}$ 
 \\
 \hline
 $VI_R$ & $u_{R}^{c1}$&
 $ \stackrel{03}{(+i)}\,\stackrel{12}{(+)}|\stackrel{56}{[+]}\,\stackrel{78}{[+]}|| 
 \stackrel{9 \;10}{(+)}\;\;\stackrel{11\;12}{[-]}\;\;\stackrel{13\;14}{[-]}$ & 
 $\nu_{R}$&
 $ \stackrel{03}{(+i)}\,\stackrel{12}{(+)}|\stackrel{56}{[+]}\,\stackrel{78}{[+]}||
 \stackrel{9 \;10}{(+)}\;\;\stackrel{11\;12}{(+)}\;\;\stackrel{13\;14}{(+)}$ 
 \\
 \hline
 $VII_R$ & $u_{R}^{c1}$&
 $\stackrel{03}{[+i]}\,\stackrel{12}{[+]}|\stackrel{56}{(+)}\,\stackrel{78}{(+)}|| 
 \stackrel{9 \;10}{(+)}\;\;\stackrel{11\;12}{[-]}\;\;\stackrel{13\;14}{[-]}$ & 
 $\nu_{R}$&
 $\stackrel{03}{[+i]}\,\stackrel{12}{[+]}|\stackrel{56}{(+)}\,\stackrel{78}{(+)}||
 \stackrel{9 \;10}{(+)}\;\;\stackrel{11\;12}{(+)}\;\;\stackrel{13\;14}{(+)}$ 
 \\
 \hline
 $VIII_R$ & $u_{R}^{c1}$&
  $ \stackrel{03}{(+i)}\,\stackrel{12}{(+)}|\stackrel{56}{(+)}\,\stackrel{78}{(+)} ||
  \stackrel{9 \;10}{(+)}\;\;\stackrel{11\;12}{[-]}\;\;\stackrel{13\;14}{[-]}$ & 
  $\nu_{R}$&
  $ \stackrel{03}{(+i)}\,\stackrel{12}{(+)}|\stackrel{56}{(+)}\,\stackrel{78}{(+)} ||
  \stackrel{9 \;10}{(+)}\;\;\stackrel{11\;12}{(+)}\;\;\stackrel{13\;14}{(+)}$ 
 \\
 \hline 
 \end{tabular}
 \end{center}
 \caption{\label{Table III.} Eight families of the right handed $u^{c1}_R$ quark with spin 
 $\frac{1}{2}$, the colour charge (${}^{c1}$) $\tau^{33}=1/2$, $\tau^{38}=1/(2\sqrt{3})$ and of 
 the colourless right handed neutrino $\nu_R$ of spin $\frac{1}{2}$ are presented in the 
 left and in the right column, respectively. All the families follow from the starting one by the 
 application of the operators ($\tilde{N}^{\pm}_{R,L}$, $\tilde{\tau}^{(2,1)\pm}$) from 
 Eq.~(\ref{plusminus}).  The generators ($N^{\pm}_{R,L}= $, $\tau^{(2,1)\pm}$) (Eq.~(\ref{plusminus}))
transform $u_{R}^{c1}$ of spin $\frac{1}{2}$ and the chosen colour ${}^{c1}$ to all the members 
of one family of the same colour. 
The same generators transform 
equivalently the right handed   neutrino $\nu_R$ of  spin $\frac{1}{2}$ to all the colourless 
members of the same family.
%the right handed  $\nu_R$ of the spin ($-\frac{1}{2}$), to  $\nu_L$ of both spins, to $e_R$ 
%of both spins and to $e_L$ of both spins. 
}
 \end{table}

 Let us point out that  the 
 break of $SO(1,7)$ into $SO(1,3) \times SU(2)_{II} \times SU(2)_{I}$,  assumed to leave 
 all the eight families massless, allows to divide eight families into two groups of four families. 
 One group of families contains  doublets with respect to $\vec{\tilde{N}}_{R}$ and 
 $\vec{\tilde{\tau}}^{2}$, these families are singlets with respect to $\vec{\tilde{N}}_{L}$ and 
 $\vec{\tilde{\tau}}^{1}$. Another group of families contains  doublets with respect to 
 $\vec{\tilde{N}}_{L}$ and  $\vec{\tilde{\tau}}^{1}$, these families are singlets with respect to 
 $\vec{\tilde{N}}_{R}$ and  $\vec{\tilde{\tau}}^{2}$. The scalar fields which are the gauge scalars 
 of  $\vec{\tilde{N}}_{R}$ and  $\vec{\tilde{\tau}}^{2}$ couple only to the four families  
 which are doublets with respect to this two groups. When gaining non zero vacuum expectation values, 
 these scalar fields determine nonzero mass matrices of the four families, to which they couple. 
 These happens at some scale, assumed that it is much higher than the electroweak scale. 
 The group of four families, which are singlets with respect to $\vec{\tilde{N}}_{R}$ and  
 $\vec{\tilde{\tau}}^{2}$, stay massless unless the gauge scalar fields of  $\vec{\tilde{N}}_{L}$ 
 and  $\vec{\tilde{\tau}}^{1}$, together with the gauge scalars of $Q\,,\,Q'\,$ and $Y'$, gain a 
 nonzero vacuum expectation values at the electroweak break. Correspondingly the decoupled twice 
 four families, that means that the matrix elements between these two groups of four families 
 are equal to zero, appear at two different scales, determined  by two different breaks.
 
To have an overview over the properties of the members of  one (any one of the eight) family let us 
present in Table~\ref{Table IV.} quantum numbers of particular members  of any of the eight families: 
The handedness 
$\Gamma^{(1+3)}(= -4i S^{03} S^{12}$), 
$\,S^{03}_{L}, \,S^{12}_L$, $\,S^{03}_{R}, \,S^{12}_R$, $\tau^{13}$ (of the weak $SU(2)_{I}$), 
$\tau^{23}$ 
(of $SU(2)_{II}$), the hyper charge $Y=\tau^{4} + \tau^{23}$, the electromagnetic charge 
$Q$, the $SU(3)$ status, 
that is, whether the member is a member of a triplet  (the quark with the one of the charges %determined by 
%$\tau^{33}$ and $\tau^{33}$ -- that is one of 
$\{(\frac{1}{2}, \frac{1}{2 \sqrt{3}}),
(-\frac{1}{2},  \frac{1}{2 \sqrt{3}} ),(0, - \frac{1}{ \sqrt{3}})\}$) or the colourless lepton, 
and $Y'$ after the break of $SU(2)_{II} \times U(1)_{II}$ into  $U(1)_{I}$. %The detailed 
%definitions of the quantum numbers in terms of $S^{ab}$ can be found in appendix~\ref{sabprop}.
%
%
 \begin{table}
 \begin{center}
 \begin{tabular}{|c||c|c|c|c|c|c|c|c|c|c|c||}
 \hline
%$i\in\{1,\cdots,8\}$ 
& $\Gamma^{(1+3)}$& $S^{03}_{L}$&$ S^{12}_L$& $S^{03}_{R}$& $S^{12}_R$& 
$\tau^{13}$ & $\tau^{23}$ & $Y$ & $Q$ & $SU(3)$&$Y'$\\
\hline
\hline
$u_{Li}$& $-1$ &$ \mp \frac{i}{2}$ &$ \pm \frac{1}{2}$& $0$& $0$& $\frac{1}{2}$& $0$&$\frac{1}{6}$& 
$\frac{2}{3}$&{\rm triplet}& $-\frac{1}{6}\, \tan^{2}\theta_{2}$\\ 
\hline
$d_{Li}$& $-1$ &$ \mp \frac{i}{2}$ &$ \pm \frac{1}{2}$& $0$& $0$& $-\frac{1}{2}$& $0$&$\frac{1}{6}$& 
$-\frac{1}{3}$&{\rm triplet}& $-\frac{1}{6}\, \tan^{2}\theta_{2}$\\
\hline
$\nu_{Li}$& $-1$ &$ \mp \frac{i}{2}$ &$ \pm \frac{1}{2}$& $0$& $0$& $\frac{1}{2}$& $0$&$-\frac{1}{2}$& 
$0$&{\rm singlet} & $\frac{1}{2}\, \tan^{2}\theta_{2}$\\
\hline
$e_{Li}$& $-1$ &$ \mp \frac{i}{2}$ &$ \pm \frac{1}{2}$& $0$& $0$& $-\frac{1}{2}$& $0$&$-\frac{1}{2}$& 
$-1$&{\rm singlet}& $\frac{1}{2}\, \tan^{2}\theta_{2}$\\
\hline\hline
$u_{Ri}$& $1$ &$0$& $0$&$\pm \frac{i}{2}$ &$ \pm \frac{1}{2}$& $0 $& $ \frac{1}{2}$&$\frac{2}{3}$& 
$\frac{2}{3}$&{\rm triplet}& $\frac{1}{2}\,(1- \frac{1}{3} \tan^{2}\theta_{2})$\\ 
\hline
$d_{R
i}$& $1$ &$0$& $0$&$\pm \frac{i}{2}$ &$ \pm \frac{1}{2}$& $0 $& $-\frac{1}{2}$&$-\frac{1}{3}$& 
$-\frac{1}{3}$&{\rm triplet} & $-\frac{1}{2}\,(1+ \frac{1}{3} \tan^{2}\theta_{2})$\\
\hline
$\nu_{Ri}$&$1$&$0$& $0$&$\pm \frac{i}{2}$ &$ \pm \frac{1}{2}$& $0 $& $ \frac{1}{2}$& $0$& 
$0$&{\rm singlet} & $\frac{1}{2}\,(1+  \tan^{2}\theta_{2})$\\
\hline
$e_{Ri}$& $1$ &$0$& $0$&$\pm \frac{i}{2}$ &$ \pm \frac{1}{2}$& $0 $& $-\frac{1}{2}$& $-1$& 
$-1$&{\rm singlet} & $-\frac{1}{2}\,(1-  \tan^{2}\theta_{2})$\\
\hline
\hline
 \end{tabular}
 \end{center}
 \caption{\label{Table IV.}  The quantum numbers of the members -- quarks and leptons, left and right handed -- 
 of any of the eight families ($i \in\{I,\cdots,VIII\}$) from Table~\ref{Table III.} are presented: The handedness 
 $\Gamma^{(1+3)}= -4i S^{03} S^{12}$, 
$S^{03}_{L}, S^{12}_L$, $S^{03}_{R}, S^{12}_R$, $\tau^{13}$ of the weak $SU(2)_{I}$, $\tau^{23}$ of the second 
$SU(2)_{II}$, the hyper charge $Y\, (=\tau^{4} + \tau^{23})$, the electromagnetic charge $Q \,(=Y + \tau^{23})$,  
the $SU(3)$ status, 
that is, whether the member is a triplet -- the quark with the one of the charges determined by 
$\tau^{33}$ and $\tau^{38}$, that is one of $\{(\frac{1}{2}, \frac{1}{2 \sqrt{3}}),
(-\frac{1}{2},  \frac{1}{2 \sqrt{3}} ),(0, - \frac{1}{ \sqrt{3}})\}$ -- or a singlet, and the charge 
$Y'\, (= {\tau^{23}- \tau^4 \,\tan^{2}\theta_{2}}$).  }
 \end{table}

Before the break of $SU(2)_{II} \times U(1)_{II}$ into $U(1)_{I}$   
the members of one family  from Tables~\ref{Table I.} and \ref{Table II.} share the  
family quantum numbers presented in 
Table~\ref{Table V.}: 
%In Table~\ref{Table V.} 
The "tilde handedness" of the families
$\tilde{\Gamma}^{(1+3)}(= -4i \tilde{S}^{03} \tilde{S}^{12})$, 
$\tilde{S}^{03}_{L}, \tilde{S}^{12}_L$, $\tilde{S}^{03}_{R}, \tilde{S}^{12}_R$ (the diagonal matrices of 
$SO(1,3)$ ), $\tilde{\tau}^{13}$ 
(of one of the two $SU(2)_{I}$), $\tilde{\tau}^{23}$ (of the second 
$SU(2)_{II}$). 
 \begin{table}
 \begin{center}
 \begin{tabular}{|l||r|r|r|r|r|r|r|r|r|r|r||}
 \hline
$i$ & $\tilde{\Gamma}^{(1+3)}$& $\tilde{S}^{03}_{L}$&$ \tilde{S}^{12}_L$& $\tilde{S}^{03}_{R}$& 
$\tilde{S}^{12}_R$& $\tilde{\tau}^{13}$ & $\tilde{\tau}^{23}$&$\tilde{\tau}^{4}$ & 
$\tilde{Y}'$&$\tilde{Y}$&$\tilde{Q}$ \\
\hline
\hline
$I$& $-1$ &$  \frac{i}{2}$ &$- \frac{1}{2}$& $0$& $0$& $ - \frac{1}{2}$& $0$&$-\frac{1}{2}$&$0$&$-\frac{1}{2}$&$-1$\\ 
\hline
$II$& $-1$ &$ -\frac{i}{2}$ &$  \frac{1}{2}$& $0$& $0$& $-\frac{1}{2}$& $0$&$-\frac{1}{2}$&$0$&$-\frac{1}{2}$&$-1$\\
\hline
$III$& $-1$ &$ \frac{i}{2}$ &$ - \frac{1}{2}$& $0$& $0$& $ \frac{1}{2}$& $0$&$-\frac{1}{2}$&$0$&$-\frac{1}{2}$&$0$\\
\hline
$IV$& $-1$ &$ -\frac{i}{2}$ &$  \frac{1}{2}$& $0$& $0$& $\frac{1}{2}$& $0$&$-\frac{1}{2}$&$0$&$-\frac{1}{2}$&$0$\\
\hline\hline
$V$& $1 $ & $0$ & $0$& $-\frac{i}{2}$ &$ -\frac{1}{2}$& $0 $& $- \frac{1}{2}$&$-\frac{1}{2}$&$\frac{1}{2}$&$-1$&$-1$\\ 
\hline
$VI\;\,$& $1 $ & $0$ & $0$& $\frac{i}{2}$ &$  \frac{1}{2}$& $0 $& $-\frac{1}{2}$&$-\frac{1}{2}$&$-\frac{1}{2}$&$-1$&$-1$\\
\hline
$VII\,$& $1 $ & $0$ & $0$& $-\frac{i}{2}$ &$- \frac{1}{2}$& $0 $& $ \frac{1}{2}$&$\frac{1}{2}$&$\frac{1}{2}$&$0$&$0$\\
\hline
$VIII$& $1 $ &  $0$& $0$& $ \frac{i}{2}$ &$\frac{1}{2}$& $0 $& $\frac{1}{2}$&$\frac{1}{2}$&$-\frac{1}{2}
$&$0$&$0$\\
\hline
\hline
 \end{tabular}
 \end{center}
 \caption{\label{Table V.}  Quantum numbers of a  member of the eight families from 
 Table~\ref{Table III.}, the same for all the members of one family, are presented: 
 The "tilde handedness" of the families 
 $\tilde{\Gamma}^{(1+3)}= -4i \tilde{S}^{03} \tilde{S}^{12}$, the left and right handed $SO(1,3)$ 
 quantum numbers (Eq.~(\ref{so13}), 
$\tilde{S}^{03}_{L}, \tilde{S}^{12}_L$, $\tilde{S}^{03}_{R}, \tilde{S}^{12}_R$ of $SO(1,3)$ group in the 
$\tilde{S}^{mn}$ sector), $\tilde{\tau}^{13}$ 
 of  $SU(2)_{I}$ , $\tilde{\tau}^{23}$ of the second 
$SU(2)_{II}$, $\tilde{\tau}^4$ (Eq.~(\ref{so6})), $\tilde{Y}'\, (= \tilde{\tau}^{23} -  
\tilde{\tau}^4 \, \tan\tilde{\theta}_2)$, taking $\tilde{\theta}^2=0$, $\tilde{Y}
\, (=\tilde{\tau}^{4} + \tilde{\tau}^{23})$, 
$\tilde{Q}= \,(\tilde{\tau}^{4} + \tilde{S}^{56})$. 
%Each member of a family carry these quantum numbers.
}
\end{table}

We see in Table~\ref{Table V.} that the first four of the eight families are singlets with respect to 
subgroups determined by  $\vec{\tilde{N}}_{R}$ and $\vec{\tilde{\tau}}^{2}$, 
and doublets with respect to $\vec{\tilde{N}}_{L}$ and $\vec{\tilde{\tau}}^{1}$, 
while the rest four families are doublets with respect to $\vec{\tilde{N}}_{R}$ and $\vec{\tilde{\tau}}^{2}$, 
and singlets with respect to $\vec{\tilde{N}}_{L}$ and $\vec{\tilde{\tau}}^{1}$.

When the  break from $SU(2)_{I} \times SU(2)_{II}\times U(1)_{II}$ to 
$\times SU(2)_{I}\times U(1)_{I}$ appears, the scalar fields, the superposition of 
$\tilde{\omega}_{abs}$, which are triplets with respect to $\vec{\tilde{N}}_{R}$ and 
$\vec{\tilde{\tau}}^{2}$ (are assumed to) gain a nonzero vacuum expectation values. 
As one can read from Eq.~(\ref{faction}) these scalar fields cause  nonzero mass 
matrices of the families which are doublets with respect to $\vec{\tilde{N}}_{R}$ 
and $\vec{\tilde{\tau}}^{2}$ and correspondingly couple to these scalar fields with nonzero 
vacuum expectation values. The four families which do not couple to these scalar fields 
stay massless. The vacuum expectation value of $\tilde{A}^{4}_{\pm}=0$ is assumed to stay 
zero at the first break.
In this break also the vector (with respect to (1+3)) gauge fields of $\vec{\tau}^{2}$ 
(the generators of $SU(2)_{II}$) become massive.

In the successive (electroweak) break the scalar gauge fields of  $\vec{\tilde{N}}_{L}$ and 
$\vec{\tilde{\tau}}^{1}$, coupled to the rest of eight families, gain  nonzero vacuum expectation values. 
Together with them also the scalar gauge fields $A^{Y'}_s$, $A^{Q'}_{s}$ and $A^{Q}_{s}$ 
(the superposition of $\omega_{sts'}$ spin connection fields) gain nonzero vacuum expectation values. 
The scalar fields $\vec{\tilde{A}}^{1}_{s}$, $\vec{\tilde{A}}^{\tilde{N}_{L}}_{s}$, $A^{Q}_{s}$, 
$A^{Q'}_{s}$ and $A^{Y'}_{s}$  determine mass matrices of the last four massless families.  
At this break also the vector gauge fields of $\vec{\tau}^{1}$ become massive.

The second break, which  (is assumed to) occurs at much lower energy scale, influences slightly also 
properties of the upper four families.

There is the contribution  which appears in the loop corrections as the term bringing 
nonzero contribution only to the mass matrix of neutrinos, transforming the right handed 
neutrinos to the left handed charged conjugated ones. It might be (by making an appropriate 
choice) that such a Majorana mass term
appears for the lower four families only. We discuss the Majorana neutrino like contribution in 
subsect.~\ref{Majoranas}.

Let us end this subsection by admitting that it is assumed (not yet showed or proved) 
that there is  no contributions to the mass matrices from 
$\psi^{\dagger}_{L} \,\gamma^0 \gamma^s \,p_{0s}\, \psi_{R},$ with 
$s=5,6$. Such a contribution to the mass term would namely mix states with different 
electromagnetic charges ($\nu_{R}$ and $e_L$, $u_{R}$ and $d_L$), in disagreement with
what is observed.

\subsection{Scalar and gauge fields in $d=(1+3)$ through  breaks}
\label{yukawandhiggs}

In the  {\it spin-charge-family} theory %as scalar fields in $d=(1+3)$-dimensional space  
%after a particular break of a symmetry %(which occurs nonadiabatically and spontaneously in a 
%kind of a phase transition) 
there are the vielbeins  $e^s{}_{\sigma} $  
\begin{displaymath}
\label{e}
 e^a{}_{\alpha} = 
\left( \begin{array}{c c} 
\delta^{m}{}_{\mu}  & 0 \\
 
 0 &  e^{s}{}_{\sigma} \\ 
\end{array}\right)
\end{displaymath}
in a strong correlation with the spin connection fields of both kinds, with 
$\tilde{\omega}_{st \sigma}$  and with $\omega_{ab \sigma} $, with indices 
$s,t, \sigma \in \{5,6,7,8\}$,  which manifest in  $d=(1+3)$-dimensional space  as scalar fields 
after  particular breaks of a  starting symmetry. Phase transitions are (assumed to be) 
triggered by the nonzero vacuum expectation values of superposition of the fields 
$f^{\alpha}{}_{s} \,
\tilde{\omega}_{ab \alpha}$ and of $f^{\alpha}{}_{s}\, \omega_{ab \alpha}$.  

The gauge fields then correspondingly appear as
\begin{displaymath}
\label{eg}
 e^a{}_{\alpha} = 
\left( \begin{array}{c c} 
\delta^{m}{}_{\mu}  & 0 \\
 e^{s}{}_{\mu}= e^{s}{}_{\sigma} E^{\sigma}{}_{Ai} A^{Ai}_{\mu}  &  \;\; e^{s}{}_{\sigma} \\ 
\end{array}\right), 
\end{displaymath}
with 
$ E^{\sigma Ai} =  \tau^{Ai} \, x^{\sigma},$ 
where $A^{Ai}_{\mu} $ are the gauge fields, corresponding to (all possible) %Kaluza-Klein 
charges $\tau^{Ai}$, manifesting in $d=(1+3)$.
Since the space symmetries  include only $S^{ab}$  ($M^{ab}= L^{ab} + S^{ab}$)  and not
$\tilde{S}^{ab}$, 
there are no vector gauge fields of the type 
$e^{s}{}_{\sigma} \tilde{E}^{\sigma}{}_{Ai} \tilde{A}^{Ai}_{\mu}$, 
with $\tilde{E}^{\sigma}{}_{Ai} =  \tilde{\tau}_{Ai} \, x^{\sigma}$.  %To the gauge fields both, 
%the vielbeins and the spin connection fields contribute~\cite{DHN,dhn}. 
The gauge fields of 
$\tilde{S}_{ab}$ manifest in $d=(1+3)$ only as scalar fields.

The vielbeins and spin connection fields from Eq.~(\ref{wholeaction}) 
($\int  d^dx \, E\; (\alpha \,R + \tilde{\alpha} \, \tilde{R})$) are  
manifesting in $d=(1+3)$ in the  effective action, if no gravity is assumed in $d=(1+3)$  
($e^m{}_{\mu}= \delta^{m}{}_{\mu} $)  
\begin{eqnarray}
\label{vbscstc2}
 S_{b} &=& \int \; d^{(1+3)}x \;\{-\frac{\varepsilon^{A}}{4}\, F^{Ai mn}\, F^{Ai}{}_{mn}  + 
 \frac{1}{2}\, (m^{Ai})^2\,\, A^{Ai}_m A^{Ai\, m} + \nonumber\\
 && {\rm contributions \; of \; scalar \; fields}\, \}.
 %\; -\frac{\tilde{\varepsilon}}{4}\, \tilde{F}^{\tilde{A}i mn}\, \tilde{F}^{ i}{}_{mn} 
 %+\nonumber\\ 
%&& \;\; (p_{0m} \Phi)^{\dagger}(p_{0}{}^{m} \Phi) - V (\Phi^{\dagger} \Phi)\},\nonumber\\
%&& \;\;  p_{0m} = p_m - \tau^{Ai} \,A^{Ai}{}_{m}, %- \tilde{\tau}^{\tilde{A}i} \, \tilde{A}^{\tilde{A}i}_{m},
\end{eqnarray}
Masses of  gauge fields of the charges $\tau^{Ai}$,  which 
symmetries are unbroken,  are zero, nonzero masses  correspond 
to the broken symmetries. 

In the breaking procedures, when  $SO(1,7) \times U(1)_{II} \times SU(3)$  breaks into 
$SO(1,3)  \times SU(2)_{I} \times SU(2)_{II} \times U(1)_{II} \times SU(3)$, there are eight massless 
families of quarks and leptons (as discussed above) and massless gauge fields 
$SU(2)_{I}, \, SU(2)_{II},\, U(1)_{II}$ and $ SU(3)$.  Gravity in $(1+3)$ is not discussed. 

In the break from $SO(1,3)  \times SU(2)_{I} \times SU(2)_{II} \times U(1)_{II} \times SU(3)\,$ to
$\,SO(1,3)  \times SU(2)_{I}  \times U(1)_{I} \times SU(3)$ the scalar fields originating in 
$f^{\alpha}{}_{s}\, \tilde{\omega}_{ab \alpha} = \tilde{\omega}_{ab s} $ gain nonzero vacuum 
expectation values causing the break of symmetries, which  manifests on the tree level 
in masses of the superposition of gauge fields $\vec{A}^{2}_{m}$ and $A^{4}_{m}$, 
as well as in mass matrices of the upper four families.

To the break from $SO(1,3)  \times SU(2)_{I}  \times U(1)_{I} \times SU(3)$ to
$SO(1,3) \times U(1) \times SU(3)$ both kinds of scalar fields, 
a superposition of $f^{\alpha}{}_{s}\, \tilde{\omega}_{ab \alpha} = \tilde{\omega}_{ab s} $ and 
 $f^{\alpha}{}_{s}\, \omega_{s't \alpha} = \omega_{s't s}, $ with 
$(s',t)= \{[5],[6],[7],[8]\}; s=\in \{[7],[8]\} $ and  $A^{4}_s$, contribute 
which manifests in the masses of $W^{\pm}_{m}, Z_{m}$ and in  mass matrices of the lower four families.  

Detailed studies of the appearance of breaks of symmetries as follow from 
the starting action, the corresponding manifestation of masses of the 
gauge fields involved in these breaks, as well as the appearance of the nonzero 
vacuum expectation values of the scalar (with respect to (1+3)) fields which manifest 
in mass matrices of the families involved in particular breaks are under 
consideration. We study in the refs.~\cite{DHN,dhn} on toy models possibilities that a 
break (such a break is in the discussed cases the one from $SO(1,13)$ to 
$SO(1,3)  \times SU(2)_{I} \times SU(2)_{II} \times U(1)_{II} \times SU(3)$,  
via $SO(1,7) \times U(1)_{II} \times SU(3)$) can end up with massless fermions. 
We found in  the ref.~\cite{DHN} for a toy model  scalar vielbein and spin connection 
fields which enable massless fermions after the break. We were not been able yet, 
even not for this toy model, to solve the problem, how do particular scalar fields 
causing a break of symmetries appear and what fermion sources are responsible for their 
appearance.

In this paper  it is  (just) assumed that there occur  nonzero vacuum expectation values 
of particular scalar fields, which then cause  breaks of particular symmetries, and change  
properties of gauge fields and  of fermion fields.

Although the symmetries of the vacuum expectation values of the scalar fields 
are known when the break of symmetries is assumed, yet their values (numbers) 
are not known. Masses and potentials determining the dynamics of these scalar fields are 
also not known, and also the way how do scalar fields contribute to masses of the gauge fields, 
on the tree and below the tree level, waits  to be studied.

Let us repeat that all the gauge vector fields %, scalar or vectors, either originating in $\omega_{abc}$ 
%or in $\tilde{\omega}_{abc}$ 
are after breaks in the adjoint representations with respect to all the groups, to which the starting 
groups break. The scalar fields, however, are in the adjoint representations with respect to the 
family groups, they are singlets with respect to $Q,Q',Y'$, while they all are, due to the break 
of the vector representations  (bi-spinor representation) of $SO(4)$, doublets with respect to 
the weak charge.

\subsubsection{Scalar and gauge fields after the break from $ SU(2)_{II} \times U(1)_{II}$ to
$ U(1)_{I}$ }
\label{scalarsubI}

Before the break of $SO(1,3) \times SU(2)_{I} \times SU(2)_{II} \times U(1) 
\times SU(3)\,$ to $\,SO(1,3)\times SU(2)_{I} \times U(1) \times SU(3)$ the gauge fields %~\cite{nproc2007}  
$\vec{A}^{2}_{m}$ ($A^{21}_m = \omega_{58m} + \omega_{67m}$, $A^{22}_m= \omega_{57m} - \omega_{68m}$,  
 $A^{23}_{m}= \omega_{56m} + \omega_{78m}$), $\vec{A}^{1}_{m}$ ($A^{11}_{m} =  
 \omega_{58m} - \omega_{67m}$, $ A^{12}_{m}= \omega_{57m}+ \omega_{68m}$, 
 $A^{13}_{m}= \omega_{56m} - \omega_{78m}$) and  $A^{4}_{m}$ %($A^{Y'}_m$ (expressible with $\omega_{stm},  
are all massless. 

After the break the gauge fields %~\cite{nproc2007} 
$A^{2\pm}_{m}$, 
as well as one superposition of $A^{23}_{m}$ and $A^{4}_{m}$,  
become massive, while  another superposition ($A^{Y}_m$) and the gauge fields $\vec{A}^{1}_{m}$ 
stay massless, due to the (assumed) break of symmetries, caused by the vacuum expectation values of 
particular superposition of the scalar fields.

The fields $A^{Y'}_{m}$  and $A^{2 \pm}_{m}$, manifesting 
as massive fields, and $A^{Y}_{m}$ which stay massless,  are defined as the superposition of the 
old ones as follows 
\begin{eqnarray}
\label{newfieldssab} 
A^{23}_{m}  &=& A^{Y}_{m} \sin \theta_2 + A^{Y'}_{m} \cos \theta_2, \nonumber\\
A^{4}_{m} &=& A^{Y}_{m} \cos \theta_2 - A^{Y'}_{m} \sin \theta_2, \nonumber\\
A^{2\pm}_m &=& \frac{1}{\sqrt{2}}(A^{21}_m \mp  i A^{22}_m),
\end{eqnarray}
for $m=0,1,2,3$ and a particular value of $\theta_2$. 
The  scalar fields $A^{Y'}_{s}$, $A^{2\pm}_{s}$, $A^{Y'}_{s}$, which do not gain in this break  
any vacuum expectation values,  stay  masses. This assumption guarantees that they 
do not contribute to masses of the lower four families on the tree level.

The corresponding operators for the new charges, which couple  these new gauge fields to  
fermions, are 
\begin{eqnarray}
\label{newoperatorssab}
Y&=& \tau^{4}+ \tau^{23}, \quad Y'= \tau^{23} - \tau^{4} \tan^{2} \theta_{2}, 
\quad \tau^{2\pm} = \tau^{21}\pm i \tau^{22}. 
\end{eqnarray}
The new coupling constants become  $g^{Y}= g^{4} \cos \theta_{2}$, $g^{Y'}= g^{2} \cos \theta_{2}$,  
while $A^{2\pm}_m $ have a coupling constant $\frac{g^2}{\sqrt{2}}$. 

The break at this stage is initiated by  the scalar fields originating in $\tilde{\omega}_{abs}$, while 
the symmetries in both sectors,  $\tilde{S}^{ab}$ and $S^{ab}$,  are broken 
simultaneously. The scalar fields $\vec{\tilde{A}}^{2}_{s}$ 
(which are the superposition of $\tilde{\omega}_{abs}$, $\tilde{A}^{21}_{s}=  \tilde{\omega}_{58s} 
+ \tilde{\omega}_{67m}$, $\tilde{A}^{22}_s= \tilde{\omega}_{57s} - \tilde{\omega}_{68s}$,  
 $\tilde{A}^{23}_{s}= \tilde{\omega}_{56s} + \tilde{\omega}_{78s}$) are among those, which 
gain a nonzero vacuum expectation values. 

We have for these  scalar fields correspondingly
\begin{eqnarray}
\label{newfieldstsab} 
\tilde{A}^{23}_{s}  &=& \tilde{A}^{\tilde{Y}}_{s} \sin \tilde{\theta}_2 + \tilde{A}^{\tilde{Y}'}_{s} 
\cos \tilde{\theta}_2, \nonumber\\
\tilde{A}^{4}_{s} &=& \tilde{A}^{\tilde{Y}}_{s} \cos \tilde{\theta}_2 - \tilde{A}^{\tilde{Y}'}_{s} 
\sin \tilde{\theta}_2, \nonumber\\
\tilde{A}^{2\pm}_s &=& \frac{1}{\sqrt{2}}(\tilde{A}^{21}_s \mp  i \tilde{A}^{22}_s),
\end{eqnarray}
for $s= [7],[8]$ and a particular value of  $\tilde{\theta}_2$ (which is indeed $0$). 
These scalar fields, having a nonzero 
vacuum expectation values, define according to Eq.~(\ref{factionI})  mass matrices of the upper 
four families, which are doublets with respect to $\vec{\tilde{\tau}}^{2}$ and $\vec{\tilde{N}}_{R}$. 

To this break and correspondingly to the mass matrices of the upper four families 
also the scalar fields, which are the gauge fields of $\vec{\tilde{N}}_{R}$ (and also couple to  
only the upper four families)  
\begin{eqnarray}
\label{tildeNR}
\vec{\tilde{A}}^{\tilde{N}_R}_{s} = ( \,\tilde{\omega}_{23s}- i \,\tilde{\omega}_{01s}\, , 
\tilde{\omega}_{31s} - i \,\tilde{\omega}_{02s}\,, \tilde{\omega}_{12s} - i \,\tilde{\omega}_{03s}\,)
\end{eqnarray}
contribute. The lower four families, which are singlets with respect to both groups, stay 
correspondingly massless.

The corresponding new  operators are then 
\begin{eqnarray}
\label{newoperatorstsab}
\tilde{Y}&=& \tilde{\tau}^{4}+ \tilde{\tau}^{23}\,, \quad \tilde{Y}'= \tilde{\tau}^{23} - 
\tilde{\tau}^{4} \tan^{2} \tilde{\theta}_{2}\,, 
\quad \tilde{\tau}^{2\pm} = \tilde{\tau}^{21}\pm i \tilde{\tau}^{22}\,, \quad 
\vec{\tilde{N}}_{R}\,. 
\end{eqnarray}
New coupling constants are correspondingly $\tilde{g}^{\tilde{Y}}= \tilde{g}^{4} \cos \tilde{\theta}_{2}$,
$\tilde{g}^{\tilde{Y}'}= \tilde{g}^{2} \cos \tilde{\theta}_{2}$,  
$\tilde{A}^{2\pm}_s $ have a coupling constant $\frac{\tilde{g}^2}{\sqrt{2}}$, and 
$\vec{\tilde{A}}^{\tilde{N}_{R}}_{s}$ $\tilde{g}^{\tilde{N}_R}$.

\subsubsection{Scalar and gauge fields after the break from $SU(2)_{I} \times U(1)_{I}$ to
$ U(1)$ }
\label{scalarsubII}

To the electroweak break, when $SO(1,3) \times SU(2)_{I} \times U(1)_{I} \times SU(3)$ breaks into 
$SO(1,3) \times U(1) \times SU(3)$,  
both kinds of the scalar spin connection fields are assumed to contribute, that is  
a superposition of $\tilde{\omega}_{abs}$, which is orthogonal to the one trigging the 
first break 
 \begin{eqnarray}
 \label{newfieldsweaktildesab}
 \tilde{A}^{13}_{s} &=& \tilde{A}_{s} \sin \tilde{\theta}_1 + 
 \tilde{Z}_{s} \cos \tilde{\theta}_1,\nonumber\\ 
 \tilde{A}^{\tilde{Y}}_{s} &=& \tilde{A}_{s} \cos \tilde{\theta}_1 -  
 \tilde{Z}_{s} \sin \tilde{\theta}_1, \nonumber\\
 \tilde{W}^{\pm}_{s} &=& \frac{1}{\sqrt{2}}(\tilde{A}^{11}_{s} \mp i  \tilde{A}^{12}_{s})\,,
 \end{eqnarray}
 and 
 \begin{eqnarray}
 \label{tildeNL}
 \vec{\tilde{A}}^{\tilde{N}_L}_{s} 
 \end{eqnarray}
 and a superposition of $\omega_{sts'}$ 
 \begin{eqnarray}
 \label{QQ'Y'}
 A^{Q}_s &=& \sin \theta_{1} \, A^{13}_{s} + \cos \theta_{1}\,A^{Y}_{s}\, ,\quad 
 A^{Q'}_s =  \cos \theta_{1} \, A^{13}_{s} - \sin \theta_{1}\,A^{Y}_{s}\, ,\nonumber\\
 A^{Y'}_{s} &=& \cos \theta_{2} \, A^{23}_{s} - \sin \theta_{2}\,A^{4}_{s}\,\,.
 \end{eqnarray}
 $ s \in ([7],[8])$. While the superposition of Eqs.(\ref{newfieldsweaktildesab}, 
 \ref{tildeNL}) couple to the lower four families only, since the lower four families are doublets 
 with respect to $\vec{\tilde{\tau}}^{1}$ and $\vec{\tilde{N}}^{\tilde{N}_{L}}$, and the upper 
 four families are singlets with respect to $\vec{\tilde{\tau}}^{1}$ and 
 $\vec{\tilde{N}}^{\tilde{N}_{L}}$, the scalar fields $A^{Q}_{s}$, $A^{Q'}_{s}$and $A^{Y'}_{s}$
 (they are a superposition of $\omega_{s'ts}; \, s',t \in ([5],\cdots,[14]); s =[7],[8]$)
 couple to all the eight families, distinguishing among the family members. 
 
 %%%%%%%%%%%%%%%5505.07.2013

Correspondingly a superposition of the vector fields $\vec{A}^{1}_m$ and $A^{4}_{m}$, 
 \begin{eqnarray}
 \label{newfieldsweaksab}
 A^{13}_{m} &=& A_{m} \sin \theta_1 + Z_{m} \cos \theta_1,\nonumber\\ 
 A^{Y}_{a}  &=& A_{m} \cos \theta_1 -  Z_{m} \sin \theta_1,\nonumber\\ 
 W^{\pm}_m &=& \frac{1}{\sqrt{2}}(A^{11}_m \mp i  A^{12}_m), 
 \end{eqnarray}
  that is $W^{\pm}_m $ and $Z_{m}$, become massive, while $A_{m} $ stays with $m=0.$
  The new   operators for charges are  
 \begin{eqnarray}
 \label{newoperatorsweaksab}
 Q  &=&  \tau^{13}+ Y = S^{56} +  \tau^{4},\nonumber\\
 Q' &=& -Y \tan^2 \theta_1 + \tau^{13}, \nonumber\\
 \tau^{1\pm}&=& \tau^{11} \pm i\tau^{12},
 \end{eqnarray}
 and the new coupling constants are correspondingly $e = 
 g^{Y} \cos \theta_1$, $g' = g^{1}\cos \theta_1$  and $\tan \theta_1 = 
 \frac{g^{Y}}{g^1} $, in agreement with the {\it standard model}. We assume, for 
 simplicity,  in the scalar sector of $\omega_{s',t,s}$ the same $\theta_{1}$ 
  as in the vector one.

 In the sector of the $\tilde{\omega}_{abs}$ scalars the corresponding new operators are  
 \begin{eqnarray}
 \label{newoperatorsweaktildesab}
 \tilde{Q}  &=&  \tilde{\tau}^{13}+ \tilde{Y} = \tilde{S}^{56} +  \tilde{\tau}^{4},\nonumber\\
 \tilde{Q'} &=& -\tilde{Y} \tan^2 \tilde{\theta}_1 + \tilde{\tau}^{13},\nonumber\\
 \tilde{\tau}^{1\pm}&=& \tilde{\tau}^{11} \pm i\tilde{\tau}^{12}, 
 \end{eqnarray}
 with the new coupling constants $\tilde{e} = 
 \tilde{g}^{Y}\cos \tilde{\theta}_1$, $\tilde{g'} = 
 \tilde{g}^{1}\cos \tilde{\theta}_1$  and $\tan \tilde{\theta}_1 = 
\frac{\tilde{g}^{Y}}{\tilde{g}^1} $. 

To this break and correspondingly to the mass matrices of the lower four families 
also the scalar fields $\vec{\tilde{A}}^{\tilde{N}_L}_{s}$, orthogonal to  
$\vec{\tilde{A}}^{\tilde{N}_R}_{s}$, contribute.

All the scalar fields are (massive) dynamical fields, coupled to fermions and governed by the 
corresponding scalar potentials, which we assume that they behave as normalizable ones (at least 
up to some reasonable accuracy).

\section{Mass matrices on the tree level and beyond}
\label{yukawatreefbelow}

In  two subsections~(\ref{scalarsubI},~\ref{scalarsubII}) of  
section~\ref{yukawandhiggs}  
properties of  scalar and gauge fields after each of two successive breaks are discussed. 
The appearance of the vacuum expectation values of some superposition of two kinds of spin 
connection fields and vielbeins, all scalars with respect to ($1+3$), is assumed which  %, 
%together with the dynamical properties of scalar and massive vector gauge fields,    
in the {\it spin-charge-family} theory  determine mass matrices of fermions and masses of vector 
gauge fields on the tree level. It is the purpose of this section  to discuss 
properties of families of  fermions fields after these two breaks,  on the tree level 
and  beyond the tree level. Properties of the family members within  the pairs  $(u\,,\nu)$ and 
$(d\,,e)$  are, namely, on the tree level very much related and it is expected that 
hopefully loop corrections (in all orders) %to the tree level mass matrices 
make  properties of the lowest three  families in agreement with the observations.

The starting fermion action (Eq.~(\ref{faction})) manifests, after the two successive 
breaks of  symmetries,  the effective low energy action presented in Eq.~(\ref{factionI}). 
The mass term (Eq.~(\ref{factionM})) manifests correspondingly in the fermion mass matrices.

Let us repeat the assumptions made to come from the starting action  to the low energy 
effective action: 
{\bf i.} In the break from $SU(2)_{I} \times SU(2)_{II} \times U(1)_{II}$ to 
$SU(2)_{I} \times U(1)_{I}$  the superposition of the $\tilde{\omega}_{abs}$
scalar fields which are the gauge  fields of $\vec{\tilde{\tau}}^{2}$ and $\vec{\tilde{N}}_{R}$, 
with the index $s \in([7]\,,[8])$
gain non zero    vacuum expectation values. 
{\bf ii.} In the electroweak break  the superposition of the  $\tilde{\omega}_{abs}$ scalar fields 
which  are the gauge fields of  $\vec{\tilde{\tau}}^{1}$ and  $\vec{\tilde{N}}_{L}$,  and 
the superposition of scalar fields $\omega_{s'ts}$ ($s',t \in([5],[6].[7],[8])$, $s=([7],[8])$), 
which are the gauge fields 
of  $Q$, $Q'$ and $Y'$, gain 
nonzero vacuum expectation values. 

The first break leaves the lower four families, which are singlets with respect to the groups 
($\vec{\tilde{\tau}}^{2}$ and  $\vec{\tilde{N}}_{R}$ ) involved in the break, massless. At the 
electroweak break all the families become massive. 
While the scalar  fields coupled with $\vec{\tilde{\tau}}^{1}$ and  $\vec{\tilde{N}}_{L}$ to 
fermions influence only the lower four families, the scalar gauge fields coupled with $Q$, $Q'$ 
and $Y'$to fermions influence  mass matrices of all the eight families.

To loop corrections the gauge vector fields, the scalar dynamical fields originating in $\omega_{s'ts}$ 
and in $\tilde{\omega}_{abs}$ contribute, those to which a particular group of families couple.
Let us tell that there is also a contribution to loop corrections, manifesting as 
a very special products of superposition of $\omega_{abs}$, %$ s=5,6,9,\cdots,14$ 
and $\tilde{\omega}_{abs}$, %$\,, \, s=5,6,7,8$ fields, 
which couple to
the right handed neutrinos and their charge conjugated states of the lower four families. This term  
might strongly influence properties of neutrinos of the lower four families.

Table~\ref{Table VII.} represents the mass matrix elements on the tree level for the upper 
four families after the first break, originating in the vacuum expectation values of 
two superposition of $\tilde{\omega}_{ab s}$ scalar fields, the two triplets of 
$\vec{\tilde{\tau}}^{2}$ and  $\vec{\tilde{N}}_R$. 
The notation $\tilde{a}^{\tilde{A}i}_{\pm}=$ $-\tilde{g}^{\tilde{A}i}\, \tilde{A}^{\tilde{A}i}_{\pm
}$ is used. The sign $(\mp)$ distinguishes  between the values of the two pairs ($u$-quarks, $\nu$-lepton)  
and ($d$-quark, $e$-lepton), respectively.  
The lower four families, which are singlets with respect to the two groups ($\vec{\tilde{\tau}}^{2}$ 
and  $\vec{\tilde{N}}_R$), as can be seen in Table~\ref{Table IV.},  stay massless after the 
first break.
 \begin{table}
 \begin{center}
\begin{tabular}{|r||c|c|c|c|c|c|c|c||}
\hline
 &$ I $&$ II $&$ III $&$ IV $&$ V $&$ VI $
 &$ VII $&$ VIII$\\
\hline\hline
$I \;\;\; $ & $ \quad 0 \quad $ & $ \quad 0 \quad $ & $ \quad 0 \quad $ & $ \quad 0 \quad$
          & $ 0 $ & $ 0 $ & $ 0 $ & $ 0 $\\
\hline
$II\;\; $ & $ 0 $ & $ 0 $ & $ 0 $ & $ 0 $
          & $ 0 $ & $ 0 $ & $ 0 $ & $ 0 $\\
\hline
$III\;$ & $ 0 $ & $ 0 $ & $ 0 $ & $ 0 $
          & $ 0 $ & $ 0 $ & $ 0 $ & $ 0 $\\ 
\hline
$IV\;\, $ & $ 0 $ & $ 0 $ & $ 0 $ & $ 0 $
          & $ 0 $ & $ 0 $ & $ 0 $ & $ 0 $\\
\hline\hline
$ V \;\; $ & $ 0 $ & $ 0 $ & $ 0 $ & $ 0 $ & 
$ -\frac{1}{2}\, (\tilde{a}^{23}_{\pm} + \tilde{a}^{\tilde{N}^{3}_{R}}_{\pm})$ & 
$ -\tilde{a}^{\tilde{N}_{R}^{-}}_{\pm} $& $0$& $  - \tilde{a}^{2-}_{\pm}$ \\
\hline
$ VI \;\,$ & $ 0 $ & $ 0 $ & $ 0 $ & $ 0 $ &
$ -\tilde{a}^{\tilde{N}_{R}^{+}}_{\pm} $&
$ \frac{1}{2}(-\tilde{a}^{23}_{\pm } + \tilde{a}^{\tilde{N}^{3}_{R}}_{\pm}) $ & 
$ -\tilde{a}^{2-}_{\pm}$ & $ 0 $ \\ 
\hline
$VII \;$   & $ 0 $ & $ 0 $ & $ 0 $ & $ 0 $ & 
$0$        & $ -\tilde{a}^{2+}_{\pm}$ &  
$  \frac{1}{2}\,( \tilde{a}^{23}_{\pm}  - \tilde{a}^{\tilde{N}^{3}_{R}}_{\pm}) $ & 
$ - \tilde{a}^{\tilde{N}_{R}^{-}}_{\pm} $  \\
\hline
$VIII$ & $ 0 $ & $ 0 $ & $ 0 $ & $ 0 $ & 
$ - \tilde{a}^{2+}_{\pm}$ & $0$& $ - \tilde{a}^{\tilde{N}_{R}^{+}}_{\pm}  $ &  
$  \frac{1}{2}\,( \tilde{a}^{23}_{\pm} + \tilde{a}^{\tilde{N}^{3}_{R}}_{\pm})$ \\
\hline\hline
\end{tabular}
 \end{center}
 \caption{\label{Table VII.}  The mass matrix for the eight families of quarks and 
 leptons after the break of $SO(1,3) \times SU(2)_{I}  \times SU(2)_{II} \times U(1)_{II} \times SU(3)$ 
 to $SO(1,3) \times  SU(2)_{I} \times U(1)_{I} \times SU(3)$. 
 %The contribution   from only the term $\tilde{S}^{ab} \,\tilde{\omega}_{abs}$ in $p_{0s}$ 
 %in Eq.(\ref{Yaction1}) 
 %is presented. 
 The notation $\tilde{a}^{\tilde{A}i}_{\pm}$ stays for 
 $-\tilde{g}^{\tilde{A}i}\, \tilde{A}^{\tilde{A}i}_{\pm}$,  $(\mp)$ distinguishes $u_{i}$ from
 $d_{i}$ and $\nu_{i}$ from $e_{i}$, index $i$ determines families.
 }
\end{table} 

Masses of the lowest of the higher four family were evaluated in the ref.~\cite{gn} from the cosmological 
and direct measurements, when assuming that baryons of this stable family (which has no mixing matrix 
elements to the lower four families) constitute the dark matter.

The lower four families obtain masses when the second $SU(2)_{I} \times U(1)_{I}$  
break occurs, at the electroweak scale, manifesting in nonzero vacuum expectation values 
of the  two triplet scalar fields $\tilde{A}^{1i}_{s}$,  $\tilde{A}^{\tilde{N}_{L}i}_{s}$, 
and the $U(1)$ scalar fields $\tilde{A}^{4}_{s}$,
as well as $A^{Q}_s$, $A^{Q'}_s$ and $A^{Y'}_s$, and also in nonzero masses of the gauge fields 
$W^{\pm}_{m}$ and $Z_m$.

Like in the case of the upper four families, also here is the mass matrix contribution from 
the nonzero vacuum expectation values of $f^{\sigma}{}_{s}\, \tilde{\omega}_{ab \sigma}$ 
on the tree level the same for ($u$-quarks and  $\nu$-leptons) and the  same for  ($d$-quarks and  
$e$-leptons), while $(\mp)$ distinguishes between 
the values of the $u$-quarks and $d$-quarks and correspondingly 
between the values of $\nu$ and $e$. The contributions from  $A^{Q}_s$, $A^{Q'}_s$ and $A^{Y'}_s$  
to mass matrices are different for different family members and the same for all the families of 
a particular family member.

Beyond the tree level all mass matrix elements of a family member become dependent on the family 
member quantum number, through  coherent contributions of the vector and all the scalar  dynamical 
fields.

Table~\ref{Table VIII.} represents the contribution of $\tilde{g}^{\tilde{1}i}\,\tilde{\tau}^{\tilde{1}i}\,
\tilde{A}^{\tilde{1}i}_{\mp}$ and $\tilde{g}^{\tilde{N}_L}\,\tilde{N}^{i}_{L}\,
\tilde{A}^{\tilde{N}_{L}i}_{\mp}$,  to the mass matrix elements on the tree level for the lower 
four families after the electroweak break. The contribution from  $e \,Q A^{Q}_s\,,$ $g^{Q'}Q' A^{Q'}_s\,$  
and $ g^{Y'}\,Y' A^{Y'}_s\,$,  which are diagonal and 
equal for all the families, but distinguish among the members of one family, are not present.  
The notation $\tilde{a}^{\tilde{A}i}_{\pm}=$ $-\tilde{g}^{\tilde{A}i}\, \tilde{A}^{\tilde{A}i}_{\mp}$ 
is used, $\tilde{\tau}^{Ai}$ stays for $\tilde{\tau}^{1i}$ and $\tilde{N}^{i}_{L}$ and correspondingly 
also the notation for the coupling constants and the triplet scalar fields is used. 
 \begin{table}
 \begin{center}
\begin{tabular}{|r||c|c|c|c||}
\hline
 &$ I $&$ II $&$ III $&$ IV $\\
\hline\hline
$I \;\;$& $ - \frac{1}{2}\, (\tilde{a}^{13}_{\pm} + \tilde{a}^{\tilde{N}^{3}_{L}}_{\pm})$ & 
$ \tilde{a}^{\tilde{N}_{L}^{-}}_{\pm} $& $0$&$   \tilde{a}^{1-}_{\pm}$ \\
\hline
$II\;$ &  $ \tilde{a}^{\tilde{N}_{L}^{+}}_{\pm}   $& 
$ \frac{1}{2}( -\tilde{a}^{13}_{\pm } + \tilde{a}^{\tilde{N}^{3}_{L}}_{\pm}) $&
$\tilde{a}^{1-}_{\pm}$& $0$\\ 
\hline
$III$ & $0$& $\tilde{a}^{1+}_{\pm}$ &
$  \frac{1}{2}\,( \tilde{a}^{13}_{\pm}  - \tilde{a}^{\tilde{N}^{3}_{L}}_{\pm}) $ & 
$ \tilde{a}^{\tilde{N}_{L}^{-}}_{\pm} $ \\
\hline
$IV$ & $  \tilde{a}^{1+}_{\pm}$ &$0$&$  \tilde{a}^{\tilde{N}_{L}^{+}}_{\pm}  $ &  
$  \frac{1}{2}\,( \tilde{a}^{13}_{\pm} + \tilde{a}^{\tilde{N}^{3}_{L}}_{\pm})$ \\
\hline\hline
\end{tabular}
 \end{center}
 \caption{\label{Table VIII.}  The mass matrix on the tree level for the lower four  families of quarks and 
 leptons after the electroweak break. Only the contributions coming  from the terms 
 $\tilde{S}^{ab} \,\tilde{\omega}_{abs}$ in $p_{0s}$ in Eq.(\ref{factionI}) are presented. 
 The notation $\tilde{a}^{\tilde{A}i}_{\pm}$ stays for $-\tilde{g}\, \tilde{A}^{\tilde{A}i}_{\pm}$, 
 where $(\mp)$  distinguishes  between the values of the ($u$-quarks and $d$-quarks) and 
between the values of ($\nu$ and $e$). 
 The terms coming from $S^{ss'}\, \omega_{ss'\,t}$ are not presented here. They are the same for 
 all the families,  but distinguish among the family members.  
 }
\end{table} 

The absolute values of the vacuum expectation values of the scalar fields contributing to the 
first break are expected to be much larger than those contributing to the second break 
($|\frac{\tilde{A}^{1i}_s}{\tilde{A}^{2i}_s}|\ll1$). 

The mass matrices of the lower four families were studied and evaluated in the ref.~\cite{gmdn} 
under the assumption 
that if going beyond the tree level the differences in the mass matrices of different 
family members start to manifest. The symmetry properties of the mass matrices from 
Table~\ref{Table VIII.}  were assumed while fitting the matrix elements to the experimental data 
for the three observed families within the accuracy of the experimental data.

In Table~\ref{Table MQN} we present quantum numbers of all members of a family, any one, 
after the electroweak break.
 \begin{table}
 \begin{center}
 \begin{tabular}{|r||r||r||r||r|||r||r||r||r||r||}
 \hline
 $$ &$Y$&$Y'$&$Q$&$Q'$& 
 $$ &$Y$&$Y'$&$Q$&$Q'$\\
 \hline \hline
 $u_R$  & $ \frac{2}{3}$& $ \frac{1}{2}\,(1-\frac{1}{3} \tan^2 \theta_2)$ & 
 $ \frac{2}{3}$& $             -\frac{2}{3} \tan^2 \theta_1 $&  
 $u_L$  & $ \frac{1}{6}$& $                -\frac{1}{6} \tan^2 \theta_2 $ & 
 $ \frac{2}{3}$& $\frac{1}{2}(1-\frac{1}{3} \tan^2 \theta_1)$
 \\
 \hline
 $d_R$  & $-\frac{1}{3}$& $-\frac{1}{2}\,(1+\frac{1}{3} \tan^2 \theta_2)$ &  
 $-\frac{1}{3}$& $              \frac{1}{3} \tan^2 \theta_1 $&
 $d_L$  & $ \frac{1}{6}$& $                -\frac{1}{6} \tan^2 \theta_2 $ & 
 $-\frac{1}{3}$& $-\frac{1}{2}(1+\frac{1}{3} \tan^2 \theta_1)$ 
 \\
 \hline
 $\nu_R$& $0$           & $\frac{1}{2}\,(1+             \tan^2 \theta_2)$ & 
 $0$&$0$&  
 $\nu_L$& $-\frac{1}{2}$&$\frac{1}{2}\,                 \tan^2 \theta_2 $ & 
 $0$& $ 0$ 
 \\
 \hline
 $e_R$ & $-1$           & $\frac{1}{2}\,(-1+             \tan^2 \theta_2)$&  
         $-1$           &                                $\tan^2 \theta_1$&  
 $e_L$ & $-\frac{1}{2}$ &$ \frac{1}{2}                  \tan^2 \theta_2 $ & 
         $-1$& $ -\frac{1}{2}(1-                          \tan^2 \theta_1)$
 \\
 \hline 
 \end{tabular}
 \end{center}
 \caption{\label{Table MQN} The quantum numbers $Y, Y',Q, Q'$ of 
 the members of a family.}
 \end{table}
It is easy to show that the contribution of complex conjugate  to $
 \psi_{L}^{\dagger}\, \gamma^{0}\, (\stackrel{78}{(-)}\,p_{\mp}\, \psi_{R}\, $  gives the same 
 value. 

Table~\ref{Table FQN} presents the quantum numbers $\tilde{\tau}^{23}$, 
$\tilde{N}^{3}_{R}\,$, $\tilde{\tau}^{13}$ and $\tilde{N}^{3}_{L}$ for all eight 
families. The first four families are singlets with respect to 
$\tilde{\tau}^{2i}$ and $\tilde{N}^{i}_{R}\,$, while they are doublets 
with respect to $\tilde{\tau}^{1i}$ and $\tilde{N}^{i}_{L}$ (all before the break 
of symmetries). The upper four families are correspondingly doublets  
with respect to $\tilde{\tau}^{2i}$ and $\tilde{N}^{i}_{R}\,$ and are singlets 
with respect to $\tilde{\tau}^{1i}$ and $\tilde{N}^{i}_{L}$.

 \begin{table}
 \begin{center}
 \begin{tabular}{|r||r||r||r||r|||r||r||r||r||r||}
 \hline
 $\Sigma=I/i$ &$\tilde{\tau}^{23}$&$\tilde{N}^{3}_{R}$&$\tilde{\tau}^{13}$ &$\tilde{N}^{3}_{L}$&
 $\Sigma=II/i$ &$\tilde{\tau}^{23}$&$\tilde{N}^{3}_{R}$&$\tilde{\tau}^{13}$ &$\tilde{N}^{3}_{L}$\\
 \hline \hline
 $1$  & $0$& $ 0$ &  $ \frac{1}{2}$&$ \frac{1}{2}$&  $1 $   & $ \frac{1}{2}$&$ \frac{1}{2}$ & $0$& $ 0$
 \\
 \hline
 $2$ & $0$& $ 0$ &  $ \frac{1}{2}$&$-\frac{1}{2}$&  $2$   & $ \frac{1}{2}$&$-\frac{1}{2}$ & $0$& $ 0$ 
 \\
 \hline
 $3$& $0$& $ 0$ &  $-\frac{1}{2}$&$-\frac{1}{2}$&  $3$  & $-\frac{1}{2}$&$-\frac{1}{2}$ & $0$& $ 0$ 
 \\
 \hline
 $4$ & $0$& $ 0$ &  $-\frac{1}{2}$&$ \frac{1}{2}$&  $4$ & $-\frac{1}{2}$&$ \frac{1}{2}$ & $0$& $ 0$
 \\
 \hline 
 \end{tabular}
 \end{center}
 \caption{\label{Table FQN} The quantum numbers $\tilde{\tau}^{23}$, 
$\tilde{N}^{3}_{R}\,$, $\tilde{\tau}^{13}$ and $\tilde{N}^{3}_{L}$ for the two groups of four 
families %($\Sigma=II$ for the upper four families and $\Sigma =I$ for the lower four families) 
are presented.}
 \end{table}

\subsection{Mass matrices  beyond the tree level}
\label{breakbelowtree}

While the mass matrices of ($u$ and $\nu$) have on the tree level  the same off diagonal elements and  
differ only in diagonal elements due to the contribution of $e \,Q A^{Q}_s\,,$ $g^{Q'}Q' A^{Q'}_s\,$ and 
$ g^{Y'}\,Y' A^{Y'}_s\,$ and the same is true for ($d$ and $e$), 
loop corrections, to which   massive gauge fields and dynamical scalar fields of both 
origins ($\tilde{\omega}_{abs}$ and $\omega_{s'ts}$) contribute coherently, are expected to change  
mass matrices of the lower four families drastically. For the upper four families, for which the 
diagonal terms from  $e \,Q A^{Q}_s\,,$ $g^{Q'}Q' A^{Q'}_s\,$ and $ g^{Y'}\,Y' A^{Y'}_s\,$  are  almost 
negligible, since they are the same for all eight families,  loop corrections are not expected to 
bring drastic changes in mass matrices between different family members.
On the tree level the mass matrices demonstrate twice four by diagonal matrices (this structure stays 
unchanged also after taking into account loop corrections in all orders)
\begin{equation} 
\label{M}
M^{\alpha} = \left(\begin{array}{cc} M^{\alpha\,II} & 0\\
0&M^{\alpha\,I} 
\end{array}\right)\,, 
\end{equation}
where $M^{\alpha\,II}_{(o)}$ and $M^{\alpha\,I}_{(o)}$ have, after taking into account loop corrections 
in all orders, the structure 
\begin{equation} 
\label{M0}
M = \left(\begin{array}{cccc} - a_1 & b
& e & c\\ b  & - a_2 & c & e\\ e & c & a_1 & b\\
c &  e & b & a_2
\end{array}\right)\,, 
\end{equation} 
with the matrix elements $a_1 \equiv a^{\Sigma}_{ \pm \,1}$, $a_2\equiv a^{\Sigma}_{\pm 2}$, 
$b \equiv b^{\Sigma}_{\pm}$, $c \equiv c^{\Sigma}_{\pm}$ and 
$e \equiv e^{\Sigma}_{\pm}$. The values $a_1\,, a_2\,,b\,,c $ and $e$ are different for the upper 
($\Sigma=II$) and the lower ($\Sigma=I$) four families, due to two different scales of  
two different breaks. One has on the tree level
\begin{eqnarray}
\label{tildea}
a_1&=& \frac{1}{2} (\tilde{a}^{(1,2)3}_{\pm} - \tilde{a}^{\tilde{N}_{(R,L)} 3}_{\pm}) \quad\, , \quad
a_2= \frac{1}{2} (\tilde{a}^{(1,2)3}_{\pm} + \tilde{a}^{\tilde{N}_{(R,L)} 3}_{\pm}) \quad \,, \nonumber\\ 
b&=&\tilde{a}^{\tilde{N}_{(R,L)} +}_{\pm}= \tilde{a}^{\tilde{N}_{(R,L)} -}_{\pm}\,, \quad
c=\tilde{a}^{(1,2)+}_{\pm}= \tilde{a}^{(1,2)-}_{\pm} \quad \,, e=0\, .
\end{eqnarray}
For the upper   four families ($\Sigma=II$) we have correspondingly 
$\tilde{a}^{3}_{\pm}= \tilde{a}^{23}_{\pm},
$ $\tilde{a}^{\tilde{N} 3}_{\pm}= \tilde{a}^{\tilde{N}_{R} 3}_{\pm}$, 
$\tilde{a}^{\pm}_{\pm}= \tilde{a}^{21}_{\pm} \pm i\, \tilde{a}^{22}_{\pm}$, 
$\tilde{a}^{\tilde{N} \pm}_{\pm}= \tilde{a}^{\tilde{N}_{R} 1}_{\pm}
\pm  i \, \tilde{a}^{\tilde{N}_{R} 2}_{\pm}$ and for the 
lower four families ($\Sigma =I\,$) we must take $\tilde{a}^{3}_{\pm}= \tilde{a}^{13}_{\pm},
$ $\tilde{a}^{\tilde{N} 3}_{\pm}= \tilde{a}^{\tilde{N}_{L}3}_{\pm}$, 
$\tilde{a}^{\pm}_{\pm}= \tilde{a}^{11}_{\pm} \pm i\, \tilde{a}^{12}_{\pm}$, 
$\tilde{a}^{\tilde{N}\pm}_{\pm}= \tilde{a}^{\tilde{N}_{L} 1}_{\pm}
\pm  i \, \tilde{a}^{\tilde{N}_{L}2}_{\pm}$.

To the tree level contributions of the scalar $\tilde{\omega}_{ab \pm}$ fields, 
diagonal matrices  $a_{\pm}$ have to be added, the same for all the eight families and different for 
each of the family member ($u,d,\nu,e$), $ (\hat{a}_{\mp} \equiv a^{\alpha}_{\mp})\,\psi$, 
which are the tree level contributions of the scalar $\omega_{sts'}$ fields
\begin{eqnarray}
\label{diagonaltreemain}
\hat{a}_{\pm}&=& e\, \hat{Q}\, A_{\pm} +  g^{1} \,\cos \theta_1\, \hat{Q}'\, Z^{Q'}_{\pm}\, +
         %\frac{g^{1}}{\sqrt{2}}\, (\tau^{1+} W^{1+}_{\pm} + \tau^{1-} W^{1-}_{\pm})
               g^{2}\, \cos \theta_2\, \hat{Y'}\, A^{Y'}_{\pm}\,.
\end{eqnarray}
Since the upper and the lower four family mass matrices appear at two
completely different scales, determined by two orthogonal sets of scalar fields, the two 
tree level mass matrices ${\cal M}^{\alpha \,\Sigma}_{(o)}$  have  very little in common, besides 
the symmetries and 
the contributions from Eq.~(\ref{diagonaltreemain}).  

%%%%%%%%%%
Let us introduce  the notation, which would help to make clear the loop corrections contributions. 
We have before the two breaks  two times ($\Sigma \in \{II,I\}$,
$II$ denoting the upper four and $I$  the lower four families) four 
massless vectors $\psi^{\alpha }_{\Sigma (L,R)}$ for each member of a  
family $\alpha \in\{u,d,\nu,e\}$. 
Let $i, \, i\in\{1,2,3,4\}\,$, denotes one of the four family members of each of the two 
groups of massless families 
\begin{equation}
\label{notationvecmassless}
\psi^{\alpha}_{\Sigma (L,R)} = \left( \psi^{\alpha}_{\Sigma\,1},\, \psi^{\alpha }_{\Sigma\,2},\,
\psi^{\alpha}_{\Sigma \,3},\, \psi^{\alpha}_{\Sigma\,4}\right)_{(L,R)}\,.
\end{equation}
Let $\Psi^{\alpha}_{\Sigma (L,R)}$ be the final massive four vectors for each of the 
two groups of families, with all loop corrections 
included
\begin{eqnarray}
\label{notatiovecnmass}
\psi^{\alpha}_{\Sigma\, (L,R)} &=& V^{\alpha}_{\Sigma} \,\Psi^{\alpha}_{\Sigma \,(L,R)} \,,\nonumber \\
V^{\alpha}_{\Sigma} &=& V^{\alpha}_{\Sigma\,(o)}\,V^{\alpha}_{\Sigma\,(1)}\,
\cdots V^{\alpha}_{\Sigma\,(k)} \cdots \,.
\end{eqnarray}
Then $\Psi^{\alpha \, (k)}_{ \Sigma (L,R)} $, which include up to $(k)$ loops corrections, read
\begin{eqnarray}
\label{notationvonek}
V^{\alpha}_{\Sigma\,(o)}\,\Psi^{\alpha \,(o)}_{ \Sigma \,(L,R)} &=&   \psi^{\alpha}_{\Sigma\, (L,R)}
\,,\nonumber\\
V^{\alpha}_{\Sigma\,(o)}\, V^{\alpha}_{\Sigma\, (1)}\,\cdots V^{\alpha}_{\Sigma \,(k)}\,
\Psi^{\alpha \,(k)}_{ \Sigma (L,R)} &=& 
\psi^{\alpha}_{\Sigma (L,R)}\,.      
\end{eqnarray}
Correspondingly we have 
\begin{eqnarray}
\label{MvPsi}
&&< \psi^{\alpha}_{\Sigma\, L}|\gamma^0 \,( M^{\alpha\,\Sigma}_{(o\, k)} + \cdots +
M^{\alpha\,\Sigma}_{(o\,1)} + M^{\alpha\,\Sigma}_{(o)})  \,
 |\psi^{\alpha}_{ \Sigma \:R}> = \nonumber\\
 &&< \Psi^{\alpha \,(k)}_{\Sigma\, L}|\gamma^0 \,(V^{\alpha}_{\Sigma\,(o)}
 \,V^{\alpha}_{\Sigma\,(1)}\,\cdots V^{\alpha}_{\Sigma\,(k)})^{\dagger}\,
 (M^{\alpha\,\Sigma}_{(o\, k)} + \cdots \nonumber\\
 &&+ M^{\alpha\,\Sigma}_{(o\,1)} + M^{\alpha\,\Sigma}_{(o)})  
 \,\,V^{\alpha}_{\Sigma\,(o)}\,V^{\alpha}_{\Sigma(1)}
 \,\cdots V^{\alpha}_{\Sigma\,(k)}\,|\Psi^{\alpha\,(k)}_{\Sigma\, R}>.
\end{eqnarray}
Let us repeat that to the loop corrections  two kinds of the scalar  
dynamical fields contribute, those originating in $\tilde{\omega}_{abs}$ ($
%\tilde{g}^{\tilde{N}_R}\, \vec{\tilde{N}}_R \vec{\tilde{A}}^{\tilde{N}_R}_{s}\,,
\tilde{g}^{\tilde{Y}'} \; \hat{\tilde{Y}}'\,\tilde{A}^{\tilde{Y}'}_{s}\;$, 
$\frac{\tilde{g}^{2}}{\sqrt{2}}\, \hat{\tilde{\tau}}^{2\pm} \,\tilde{A}^{2\pm}_{s}, \;	
\tilde{g}^{\tilde{N}_{L,R}} \hat{\vec{\tilde{N}}}_{L,R}\, \vec{\tilde{A}}^{\tilde{N}_{L,R}}_{s}\,$, $
\tilde{g}^{\tilde{Q}'} \,  \hat{\tilde{Q}}'\; \tilde{A}^{\tilde{Q}'}_{s}$, 
$ \, \frac{\tilde{g}^{1}}{\sqrt{2}}\, \hat{\tilde{\tau}}^{1\pm} \,\tilde{A}^{1\pm}_{s}\,$) 
and those originating in $\omega_{abs}$   
	          ($e\, \hat{Q}\, A_{s}\,, \;g^{1}\, \cos \theta_1 \,\hat{Q}'\, Z^{Q'}_{s}\,$,  
              $ g^{Y'} \cos \theta_2\, \hat{Y}'\, A^{Y'}_{s}$) 
and the  massive gauge fields ($
               g^{2} \cos \theta_2 \,\hat{Y}'\, A^{Y'}_{m}$, $g^{1}\, \cos \theta_1 \,\hat{Q}'\, Z^{Q'}_{m}\,$)  
               as it follow from~Eq.(\ref{factionI}).           
   
In the ref.~\cite{albinonorma} the loop diagrams   for these contributions to loop corrections 
are presented and  numerical results discussed for both groups of four families. The masses 
and coupling constants of dynamical scalar fields and of the massive vector fields are taken as an 
input and the influence of loop 
corrections on properties of fermions studied.

Let us arrange  mass matrices, after the electroweak break and when  all the loop 
corrections are taken into account, as a sum of matrices as follows   
\begin{eqnarray}
\label{qprime}
M^{\alpha\, \Sigma}&=& \sum_{k=0, k'=0, k''=0}^{\infty} \,(Q^{\alpha})^{k}\, ({Q'}^{\alpha})^{k'} 
\,({Y'}^{\alpha})^{k''} \, 
 M^{\alpha\, \Sigma}_{Q \,Q'\, Y'\,k\, k'\,k''}\,. 
\end{eqnarray}
To each family member there corresponds its own matrix $M^{\alpha\, \Sigma}$. 
It is a hope, however, that the matrices $M^{\alpha\, \Sigma}_{Q \,Q'\, Y'\,k\, k'\,k''}$ might  
depend only slightly on the family member index $\alpha$ ($M^{\alpha \,\Sigma}_{Q \,Q'\, Y'\,k\, k'\,k''}= $
$M^{\Sigma}_{Q \,Q'\, Y'\,k\, k'\,k''}$) and that  the eigenvalues of the operators 
$(\hat{Q}^{\alpha})^{k}\, (\hat{Q}'^{\alpha})^{k'} \,(\hat{Y}'^{\alpha})^{k''}$
on the massless states $\psi^{\alpha}_{\Sigma\, R}$  make the mass matrices  $M^{\alpha\, \Sigma}$
dependent on $\alpha$. To  masses of neutrinos only the terms $(Q^{\alpha})^{0}\, ({Q'}^{\alpha})^{0} 
\,({Y'}^{\alpha})^{k''} \,  M^{\alpha\, \Sigma}_{Q \,Q'\, Y'\,0\, 0\,k''}$  contribute.

There is an additional term, however, which does not really speak for the suggestion of Eq.~(\ref{qprime}). 
Namely the term in loop corrections which transforms the 
right handed neutrinos into their left handed charged conjugated ones and which 
manifests accordingly the  Majorana neutrinos. These contribution is presented  in 
Eq.~(\ref{CnuRdaggernuR}) of the subsection \ref{Majoranas}. It might concern only the lower 
groups of four families and might contribute a lot, in addition to the ''Dirac masses'' of 
Eq.~(\ref{qprime}), to the extremely small masses 
of the observed families of neutrinos. This term needs, as also all the loop corrections to the tree 
level mass matrices for all the family members, additional studies.

More about the mass matrices below the tree level can be found in the 
ref.~\cite{albinonorma}.

\subsubsection{Majorana mass terms}
\label{Majoranas}

There are mass terms  within the {\it spin-charge-family} theory, which transform
the right handed neutrino to his  charged conjugated
one, contributing to the right handed neutrino Majorana masses 
\begin{eqnarray}
\label{CnuRdaggernuR}
&&\psi^{\dagger}\, \gamma^0 \stackrel{78}{(-)} p_{0-}\, \psi\,,\nonumber\\
&&p_{0-}=  - (\tilde{\tau}^{1 +} \,\tilde{A}^{1 +}_{-}+ \tilde{\tau}^{1 -} \,\tilde{A}^{1 -}_{-})\;
 {\cal O}^{[+]} \, \cal{A}^{O}_{[+]},\nonumber\\
&&{\cal O}^{[+]}\, =   \stackrel{78}{[+]}\,
\stackrel{56}{(-)}\,\stackrel{9\,10}{(-)}\,
\stackrel{11 \,12}{(-)}\;\;\stackrel{13\,14}{(-)}.
\end{eqnarray}
One easily checks, using the technique with the Clifford objects,  
that $\gamma^0 \stackrel{78}{(-)} p_{0-}$  transforms a {\it right handed neutrino} of one of the
{\it lower four families} into the charged conjugated one, belonging to the same group of families. It does not
contribute to masses of other leptons and quarks, right or left handed. Although the operator ${\cal O}^{[+]}$
appears in a quite complicated way, that is in the higher order corrections, yet it might be helpful when 
explaining the properties of neutrinos. The operator
$- (\tilde{\tau}^{1 +} \,\tilde{A}^{1 +}_{-}+ \tilde{\tau}^{1 -} \,\tilde{A}^{1 -}_{-})\;
 {\cal O}^{[+]} \, \cal{A}^{O}_{[+]}$ gives zero, when applied on the upper
four families, since the upper four families  are singlets with respect to $\tilde{\tau}^{1 \pm}$.

This term needs further studies.

\section{Scalar fields of the {\em spin-charge-family} theory  manifesting effectively as the
 {\em standard model Higgs}}
\label{scalarsHM}

Let us try to understand the {\it standard model} as an effective approach of 
the {\it spin-charge-family} theory. (The inclusion of the right handed neutrino as a regular family 
member in the {\it spin-charge-family} theory 
 is a little extension to the {\it standard model}.) 

Mass matrices of fermions of the lower four families 
are in the {\it spin-charge-family} theory, according to Eq.(\ref{factionI}), on the tree 
level determined by the scalar fields through the operator
\begin{eqnarray}
 \hat{\Phi}^{I}_{\mp}  &=&   \, \stackrel{78}{(\mp)} \,
	       \{ \tilde{g}^{\tilde{N}_L} \,  \vec{\tilde{N}}_L \, \vec{\tilde{A}}^{\tilde{N}_L }_{\mp}
	      +  \tilde{g}^{1}\, \vec{\tilde{\tau}}^{1}
		  	       	          \,\vec{\tilde{A}}^{1}_{\mp}  	
	         +   e\, Q\, A^{Q}_{\mp} +  g^{Q'}\, Q'\, Z^{Q'}_{\mp} +
              \,g^{Y'} \, Y'\, A^{Y'}_{\mp} \}\,.
\label{factionIH}
\end{eqnarray}
The operator $\stackrel{78}{(\mp)} $,  included in $\hat{\Phi}^{I}_{\mp}$, transforms 
all the quantum numbers of the right handed quarks and leptons to those of the left handed ones, 
except the handedness, % and correspondingly the spin properties in $d=(1+3)$, 
for the transformation of which in the {\it standard model} as well as in the 
{\it spin-charge-family} theory $\gamma^0$ takes care. The product of the operators $\gamma^0$ 
and $\stackrel{78}{(\mp)} $ transforms
the  right handed quarks and leptons into the left handed ones, as explained in section~\ref{actiontosm} 
and can be read in Tables~\ref{Table I.}~and~\ref{Table II.}. 
The part of the operator $\hat{\Phi}^{I}_{\mp} $, that is 
$\{ \tilde{g}^{\tilde{N}_L} \,  \vec{\tilde{N}}_L \, \vec{\tilde{A}}^{\tilde{N}_L }_{\mp} 
+  \tilde{g}^{1}\, \vec{\tilde{\tau}}^{1}  \,\vec{\tilde{A}}^{1}_{\mp} 
+   e\, Q\, A^{Q}_{\mp} +  g^{Q'}\, Q'\, Z^{Q'}_{\mp} + 
\,g^{Y'} \, Y'\, A^{Y'}_{\mp} \}\,$, takes care of the mass matrices of quarks and leptons. 
The application of  $\{ \tilde{g}^{\tilde{N}_R} \,  \vec{\tilde{N}}_R \, 
\vec{\tilde{A}}^{\tilde{N}_R}_{\mp} +  \tilde{g}^{2}\, \vec{\tilde{\tau}}^{2} \,\tilde{A}^{2}_{\mp}\,\}$ 
on the lower four families is zero, since the lower four families  are singlets with respect to 
$\vec{\tilde{N}}_R\,$ 
and $\vec{\tilde{\tau}}^{2}$. In  the loop corrections besides the massive scalar fields - 
$\vec{\tilde{A}}^{1}_{\mp}$, $\vec{\tilde{A}}^{\tilde{N}_L}_{\mp}$, $A^{Q}_{\mp}$, $Z^{Q'}_{\mp}$ and 
$A^{Y'}_{\mp}$ - also the massive gauge vector fields - $Z^{Q'}_{m}$, $A^{1 \pm}_{m}= W^{\pm}_{m}$, 
$A^{Y'}_{m}$ and $A^{2 \pm}_{m}$ - start to contribute coherently. 

Although on the tree level the contributions of the massive scalar fields to mass matrices of 
different family members are very much correlated, and correspondingly in disagreement with the 
experimental data, the loop corrections, bringing strong dependence on the family member quantum 
number $\alpha$, change mass matrices. First rough evaluations of the loop corrections 
contributions to the tree level mass matrices are giving a hope that with the loop corrections 
taken into account mass matrices reproduce measurable masses and mixing matrices.

We do not yet know the properties of the scalar fields: their vacuum expectation values, 
masses and coupling constants. We may expect that they behave similarly as the Higgs field in the 
{\it standard model}, that is that their dynamics is determined by potentials which 
make  contributions of the scalar fields renormalizable~\footnote{The proprties of the scalar fields 
are under consideration. One can see that they are triplets with respect to the family groups, 
singlets with respect to $Q,Q'$ and $Y'$ and doublets with respect to the two $SU(2)_{II}$ 
and $SU(2)_{I}$ group}. Starting from the spin connections and 
vielbeins we only can hope that at least effectively at the low energy regime, that is in the weak 
field regime, the effective theory is behaving as a renormalizable one. 

Yet we can  estimate masses of gauge bosons under the assumption that the scalar fields, 
which determine mass matrices of fermion family members, are determining effectively also 
masses of gauge fields, like this is assumed in the {\it standard 
model}. If breaking of symmetries occurs in both sectors in a correlated way (I have 
assumed so far that this is the case) then symmetries of the vielbeins $f^{\alpha}{}_{a}\,$ 
and the two spin connection fields, $\omega_{abc}= f^{\alpha}{}_{c}\, \omega_{ab\alpha}$ 
and $\tilde{\omega}_{abc} = f^{\alpha}{}_{c} \,\tilde{\omega}_{ab \alpha}$,   change 
simultaneously.

To see the {\it standard model} as an effective theory of the {\it spin-charge-family} theory,
let us  assume the existence of the scalar fields, which by "dressing" right handed family members, 
ensure them the weak and the hypercharge of their left handed partners. 
Therefore, we  replace 
the part  $\stackrel{78}{(-)} $  of the operator $\hat{\Phi_{\mp}}$ in 
Eq.(\ref{factionIH}), which transforms the weak chargeless  
right handed $u_{R}$ quark of a particular hypercharge $Y$ ($\frac{2}{3}$) of any family into the weak charged 
$u_R$ quark with $Y $ of the left handed $u_{L}$ ($\frac{1}{6}$), while $\gamma^0$ changes its 
right handedness into the left one,  %and equivalently for other family members, 
with the scalar field of Table~\ref{Table SMH.}, which has the appropriate weak and hyper charge. 
Although all the scalar fields %as all the gauge fields of the {\it spin-charge-family} theory 
are bosons, manifesting their family charges  
in the adjoint representations, they all are doublets with respect to the two $SU(2)$ charges, 
that is also with respect to the weak charge. 
Correspondingly we replace, in order to mimic the {\it standard model},  the part  
$\stackrel{78}{(-)} $ by the scalar  field with the charges originating in $S^{ab}$ in the fundamental 
representation. This scalar field is a 
colourless weak doublet with the hyper charges $Y=- \frac{1}{2}$ for ($u_R$ and $\nu_{R}$) and 
$Y=  \frac{1}{2}$ for ($(d_R)$ and $e_R$),  while only the components 
with the electromagnetic charge $Q= \tau^{13} + Y$ equal to zero are allowed to have nonzero 
vacuum expectation values. 
This massive dynamical scalar fields  with  nonzero vacuum 
expectation values  are assumed in this stage to be governed by a hopefully renormalizable potential. 
This means that effectively 
 all the scalar fields appearing in Eq.~(\ref{factionIH}) manifest at the so far observed energy regim
 as by the {\it standard model} assumed Higgs  and the Yukawa couplings.

We can simulate the part  $\stackrel{78}{(-)} $ with the scalar field 
$\Phi^{I}_{-}$, presented in Table~\ref{Table SMH.}, which "dresses" $u_{R}$  and $\nu_{R}$ 
in the way assumed by the {\it standard model}, and we simulate the part  $\stackrel{78}{(+)} $ 
with  the scalar field $\Phi^{I}_{+}$  from 
Table~\ref{Table SMH.}, which "dresses" $d_{R}$ and $e_{R}$.  
 \begin{table}
 \begin{center}
 \begin{tabular}{|r||c||r|r|r|r|r||c||}
 \hline\hline
&$\Phi^{I}$ & $\tau^{13}$& $\tau^{23}$& $\tau^{4}$&$ Y$ & $Q$& colour charge\\
\hline
$ \Phi^{I}_{-}$& $\stackrel{56}{[+]}\,\stackrel{78}{[-]}
|| \stackrel{9 \;10}{[+]}\;\;\stackrel{11\;12}{[+]}\;\;\stackrel{13\;14}{[+]}$
&$ \frac{1}{2}$ &$0$&$ -\frac{1}{2}$&$-\frac{1}{2}$& $0$& colourless\\
\hline
$ {\bf \Phi^{I}_{+}}$& ${\bf \stackrel{56}{[-]}\,\stackrel{78}{[+]}}  
||  \stackrel{9 \;10}{[-]}\;\;\stackrel{11\;12}{[-]}\;\;\stackrel{13\;14}{[-]}$
&$-\frac{1}{2}$ &$0$&$  \frac{1}{2}$&$ \frac{1}{2}$&$0$& colourless\\
\hline\hline
 \end{tabular}
 \end{center}
 \caption{\label{Table SMH.}  One possible choice of  the weak and hypercharge components of 
 the  scalar fields carrying the quantum numbers of the  {\it standard model} Higgs, 
 presented in the technique~\cite{norma,snmb:hn02hn03}, chosen to play the role of the 
 {\it standard model} Higgs. Both states on the table are colour singlets, 
 with the weak and hypercharge, which if used in the {\it standard model} way "dress" 
 the right handed quarks and leptons so that they carry  
 quantum numbers of the left handed partners.  
 The state ($\Phi^{I}_{-} \, u_R$), for example, carries the weak and the hypercharge of $u_L$. 
  "Dressing" the right handed family members with the $\Phi^{I}_{\mp}$ manifests 
 effectively  as the  application of the operators $\stackrel{78}{(\mp)}$  (Eq.(\ref{factionIH}))  
 on the right handed family members. 
 }
\end{table}

In the {\it spin-charge-family} theory  are the mass matrices  after the electroweak break 
determined on the tree level by the expectation values of the operator 
%$\{ \tilde{g}^{\tilde{N}_L} \,  \tilde{N}^{i}_L \, \tilde{A}^{\tilde{N}_L \,i}_{\mp} 
%+  \tilde{g}^{1}\, \tilde{\tau}^{1i} \,\tilde{A}^{1i}_{\mp} +   e\, Q\, A^{Q}_{\mp} +  
%g^{Q'}\, Q'\, Z^{Q'}_{\mp} +  \,g^{Y'} \, Y'\, A^{Y'}_{\mp} \}$  (Eq.(\ref{factionIH})) 
%for each particular member. %Let  this operator  be called 
$\hat{\Phi}^{vI}_{\mp}  $   (Eq.(\ref{factionIH})) 
\begin{eqnarray}
 \hat{\Phi}^{vI}_{\mp}  &=&   \,
	       \{ \tilde{g}^{\tilde{N}_L} \,  \tilde{N}^{i}_L \, \tilde{A}^{\tilde{N}_L \,i}_{\mp}
	      +  \tilde{g}^{1}\, \tilde{\tau}^{1i}
		  	       	          \,\tilde{A}^{1i}_{\mp} \nonumber\\ 	
	         &+&   e\, Q\, A^{Q}_{\mp} +  g^{Q'}\, Q'\, Z^{Q'}_{\mp} +
              \,g^{Y'} \, Y'\, A^{Y'}_{\mp} \}\,. 
\label{factionIHpart}
\end{eqnarray}
%
%$\hat{\Phi}^{vI}_{\mp} $ 
In the attempt to see the {\it standard model} as an effective theory of the  
{\it spin-charge-family} theory    the {\it standard model} Higgs together with the  
Yukawa couplings can  be understood as the interaction of family members with these scalar fields,
which obviously offer to the right (left) handed family members a needed weak and hyper charges 
(formally through the operator $\stackrel{78}{(\mp)} $), changing at the same time family properties
(through  $\Phi^{I}_{\mp}$). 

Masses of the vector gauge fields as well as the properties of the scalar fields  
should in the {\it spin-charge-family} theory be determined by studying  
the break of symmetries. We discuss  in subsect.~\ref{yukawandhiggs} the break, but the detailed 
calculations are very demanding and we have not  (yet) been able to perform them.   

One can extract some information about properties of the scalar fields in Eq.(\ref{factionIHpart}) 
from the  masses of  the so far observed quarks and leptons and the weak boson masses. 
From the covariant momentum after the electroweak break 
\begin{eqnarray}
\label{covp0}
p_{0m} &=& p_m - \frac{g^1}{\sqrt{2}} [\tau^{1+}\, A^{1+}_{m} +
\tau^{1-}\, A^{1-}_{m}] + g^1 \sin \theta_1\, Q \, A^{Q}_{m} + g^1 \cos \theta_1\,Q'\, A^{Q'}_{m}\,,
\end{eqnarray}
with $\theta_1$ equal $ \theta_{W}$, with the electromagnetic coupling constant 
$e= \sin \theta_W$,  the charge operators 
$Q= \tau^{13} + Y$,  $Q'= \tau^{13} - \tan^2 \theta_W \, Y$, and with the gauge fields 
$W^{\pm}_{m}= A^{1\pm}_{m} =  \frac{1}{\sqrt{2}} \, 
(A^{11}_{m} \mp A^{12}_{m}) \,$, $A_{m} = A^{Q}_{m} = A^{13}_{m} \sin \theta_W +  
A^{Y}_{m} \cos \theta_W$  and $Z_{m}= A^{Q'}_{m} = A^{13}_{m} \cos \theta_W -  
A^{Y}_{m} \sin \theta_W$,  we estimate 
\begin{eqnarray}
 (p_{0m}\, \hat{\Phi}^{I}_{\mp} )^{\dagger}\,(p_{0}^{\,\,m}\,\hat{\Phi}^{I}_{\mp})  
 &\leftarrow &   \{\,\frac{(g^1)^2}{2}\,A^{1+}_{m}\, A^{1-\,m} + (\frac{g^1}{ 2\,\cos \theta_1})^2
 \,A^{Q'}_{m}\, A^{Q'\,m}\,\} \, Tr (\Phi^{v  I\dagger}_{\mp}\,\Phi^{v I}_{\mp} )\,. 
\label{gaugemIH}
\end{eqnarray}
$\Phi^{v I}_{\mp}$ are determined in Eq.~(\ref{factionIHpart}), while the 
states $\Phi^{I}_{\mp}$  from Table~\ref{Table SMH.}  
are normalized to unity as explained in the refs.~\cite{snmb:hn02hn03} 
and in the appendix. 
Assuming, like in the {\it standard model},  that $Tr (\Phi^{v  I\dagger}_{\mp}\,\Phi^{v I}_{\mp} ) =
\frac{v^2}{2}$, we extract from the masses of gauge bosons one information about the vacuum 
expectation values of the scalar fields, their coupling constants and their masses. 
Mass matrices of quarks and leptons 
offer additional information about the scalar fields of the {\it spin-charge-family} theory.
Measuring  charged and neutral currents,  decay rates of hadrons, 
the scalar fields productions in the fermion scattering events and their decay properties 
provides us with additional in formations.

Studying neutral and charged currents and possible scalar field productions and decays are important 
next step to be done.

%Let me conclude this section by the observation that the  colourless scalar 
%with the weak charges in 
%the fundamental representation of the $SU(2)$ group is a strange object from the point of 
%view of the fact that all the known fields are either fermions in the fundamental 
%representations with respect to the charge groups or they are  (vector) bosons in the 
%adjoint representations with respect to the charge groups. The {\it standard model} Higgs 
%is therefore a very artificial object. This is not the case in the {\it 
%spin-charge-family} theory, where the scalar fields carry all the charges (with the family 
%quantum numbers included) in the adjoint representation.

%
\section{Models with the $SU(3)$ flavour groups and  the {\em spin-charge-family} theory}
\label{HolgerGuidiceGavela}
%
%%%%%%

In section~\ref{scalarsHM} we  look at the  {\it standard model} assumptions from the 
point of view of the {\it spin-charge-family} theory. 
There are many attempts in the literature to connect families of quarks and leptons with the 
fundamental representations of the $SU(3)$ gauge group. Let me comment the 
assumptions presented in the refs.~\cite{Georgi,giudice,belen} from the point of view of 
the {\it spin-charge-family} theory.

The authors of the refs.~\cite{Georgi,giudice} assumed that   the three  so far observed families 
of  quarks and leptons, if all 
 massless, manifest the "flavour" symmetry
\begin{eqnarray}
\label{mfgroup}
SU(3)_{q_{L}} \times SU(3)_{u_{R}} \times SU(3)_{d_{R}}  \times 
SU(3)_{l_{L}} \times SU(3)_{\nu_{R}} \times SU(3)_{e_{R}} \,.  
\end{eqnarray}
Each family member is a member of a $SU(3)$ triplet. %, which is different for different family members, 
%only the weak doublet carries the same family group. 
Neutrinos are not included in the original papers.
In Eq.~(\ref{mfgroup}) a small extension of the {\it standard model} is  assumed by taking
the right handed neutrino as a regular family member. The family (flavour) group of Eq.~(\ref{mfgroup}) 
distinguishes among the colour triplet  quarks  which are left handed weak doublets  with  the 
hypercharge $\frac{1}{6}$, the right handed weak singlet quarks ($u_R,d_R$) with the hypercharge 
($\frac{2}{3}, -\frac{1}{3}$), the colour singlet leptons which are left handed weak doublets 
 with  the hypercharge $-\frac{1}{2}$ and the right handed weak singlet leptons ($\nu_R,e_R$) 
 with the hypercharge ($0,-1$).

In the refs.~\cite{giudice,belen} is assumed  in addition that Yukawa scalar fields, which together 
with the vacuum expectation values of the 
Higgs fields, break the flavour symmetry of Eq.(\ref{mfgroup}), are singlets with respect to all 
the {\it standard model} gauge group while they are in the bi-fundamental representation of 
the group $SU(3)$: They are either triplets with respect to the flavour group of the left handed 
weak doublet quarks and antitriplets with respect to each of the two flavour groups 
of the right handed weak chargeless quarks, or they are triplets with respect to the flavour group 
of the left handed weak doublet leptons and antitriplets with respect to each of the two flavour groups 
of the right handed weak chargeless leptons.

In the {\it spin-charge-family} theory the families appear as  representations of the group 
 with the infinitesimal generators $\tilde{S}^{ab}$, forming the equivalent representations with respect to  
 the group with the generators $S^{ab}$. These latter generators determine spin of fermions and their charges. 
 The fields gauging the group $S^{ab}$  determine after the break of symmetries in the low energy regime 
 all the known gauge fields, which are vectors in ($1+3$) with the charges in the adjoint representations. 
 
 The scalar fields which are gauge fields of the $\tilde{\tau}^{1i}$ ($= c^{1i}{}_{ab}\, \tilde{S}^{ab}$) 
 and $\tilde{N}^{i}_{L}$ ($= c^{N_{L}i}{}_{ab}\, \tilde{S}^{ab}$)  in the adjoint representations 
 of these two $SU(2)$ groups determine together with the singlet scalar fields  
 which are the gauge  fields of $Q,Q'$ and $Y'$ (all three  are expressible with $S^{ab}$) 
 after the electroweak break  
 the mass matrices of four families of 
 quarks and leptons,  and correspondingly the Yukawa couplings, masses and mixing matrices of the
 (lowest) four families  of  quarks and leptons. The scalar fields -- which all are doublets with respect to the 
 weak charge --  determine with their nonzero 
 vacuum expectation values on the tree level mass matrices of quarks and leptons and 
 contribute to masses of weak bosons. Below the tree level the dynamical 
 scalar fields of both origins and the massive gauge fields bring coherent contributions 
 to the tree level mass matrices. 
 
 While on the tree level the off diagonal matrix elements of the  $u$-quark mass matrix are equal to the 
 off diagonal matrix elements of  the $\nu$-lepton mass matrix, and 
 the off diagonal matrix elements of the  $d$-quark mass matrix  equal to the 
 off diagonal matrix elements of  the $e$-lepton mass matrix, the loop corrections change this picture 
 drastically, hopefully reproducing the experimentally observed properties of fermions.

 All the fermion charges, with the family quantum number included,
 are described by the fundamental representations of the corresponding groups, and 
 all the bosonic  vector fields have their charges in the adjoint representations. The scalar fields,
 however, which before the break of the $SO(4)$ are vectors in ($[5],[6],[7],[8]$), are after the break 
 of $SO(4)\times U(1)_{II}$ into $SU(2)_{I}\times U(1)_{I}$ weak doublets, while they have all the 
 family charges in the  adjoint representations.

 The refs.~\cite{Georgi,giudice,belen} try to explain the appearance of mass matrices of 
 the {\it standard model}, which manifest in the Higgs fields and the Yukawa matrices,  
 by taking the Yukawas as dynamical fields. Yukawa scalar fields with their bi-fundamental 
 representations of the $SU(3)$ flavour  group are an attempt to continue with the assumption of the 
 {\it standard model} that there exist  scalar fields with charges in the fundamental 
 representations. To do the job Yukawa scalar field are assumed to be in the bi-fundamental 
 representations. %This is close to the adjoint $SU(3)$ representation. 
 It seems a very nontrivial task to 
 make use of the analyses of the experimental data  presented as a general extension of the 
 {\it standard model} in the refs.~\cite{giudice,belen} for the 
  {\it spin-charge-family} theory as it manifests in the low energy region.

% (****To poglavje preberi se enkrat in popravi  

%
\section{Conclusions}
\label{conclusions}

The  {\it spin-charge-family} theory~\cite{norma,pikanorma,Portoroz03} is offering 
the way beyond the {\it standard model}    
by proposing the mechanism for generating families of quarks and leptons and consequently  
predicting  the number of families at low (sooner or later) observable energies and the   
mass matrices for each of the family member (and correspondingly the masses and the mixing 
matrices of families of quarks and leptons).  

The  {\it spin-charge-family} theory predicts the fourth family to be possibly measured at 
the LHC %or at some higher energies 
and the fifth family which is, since it is decoupled in 
the mixing matrices from the lower four families and it is correspondingly stable, the 
candidate to form the dark matter~\cite{gn}. 

The proposed theory also predicts  that there are several scalar fields, taking care of 
mass matrices of the two times four families and of the masses of weak gauge bosons. 
At low energies these scalar dynamical fields manifest effectively pretty much 
as the {\it standard model} Higgs field together with Yukawa couplings, 
predicting at the same time that observation of these scalar fields is expected to deviate 
from what for the Higgs the {\it standard model} predicts.

To the mass matrices of fermions two kinds of scalar fields contribute, the one interacting with 
fermions through the  Dirac spin and the one interacting with fermions through 
the second kind of the Clifford operators (anticommuting with the Dirac ones). The first one 
distinguishes among the family members, the second one among the families. Beyond the tree level 
these two kinds of scalar fields and the vector massive fields start to contribute coherently, 
leading hopefully to the measured properties of the so far observed three families of fermions 
and to the observed weak gauge fields.

In the ref.~\cite{pikanorma,gmdn} we made a rough estimation of properties of quarks and 
leptons of the lower four families as predicted by the  {\it spin-charge-family} theory.  
The mass matrices of quarks and leptons turns out to be strongly related on the tree level. 
Assuming  that loop corrections change elements of mass matrices considerably, but  
keep the symmetry of mass matrices, we took mass matrix element of the lower four families as free 
parameters. 
We fitted the matrix elements to the existing experimental data for the observed three families 
within the experimental accuracy and for a chosen mass of each of the fourth family member. 
We predict then  elements of the mixing matrices for the fourth family members as well as the 
weakly measured matrix elements of the three observed families. 

In the ref.~\cite{gn} we evaluated the masses of the stable fifth family  (belonging to  the upper 
four families) under the assumption that  neutrons and neutrinos of this stable fifth family 
form the dark matter. 
We study the properties of the fifth family  neutrons, their freezing out of  the cosmic plasma 
during the evolution of the universe, as well as  their interaction among themselves and 
with the ordinary matter in the direct experiments.

In this paper we study properties of the gauge vector and scalar fields  and their influence on 
the properties of eight families of quarks and leptons as they  follow from the  
{\it spin-charge-family} theory on the tree and  below the tree level  after the  two successive 
breaks,  from $SO(1,3) \times SU(2)_{I} \times SU(2)_{II} \times U(1)_{II} \times SU(3)$ to 
$SO(1,3) \times SU(2)_{I} \times U(1)_{I} \times SU(3)$ and further to $SO(1,3) \times U(1) 
\times SU(3)$, trying to understand better what happens during these two breaks and after them.  

We made several assumptions, since we are not (yet) able to  evaluate how do these two  breaks 
occur and what does trigger them. 

In the first break several scalar %(with respect to $SO(1,3)$) 
fields (the superposition of 
$f^{\sigma}{}_{s}\, \tilde{\omega}_{abs}\,, s=([7],[8]) $) contribute to the break as gauge 
triplet fields of $\vec{\tilde{N}}_{R}$ and of $\vec{\tilde{\tau}}^{2}$, gaining nonzero 
vacuum expectation values. 
Correspondingly they cause nonzero mass matrices of the upper four families to which they couple  
and nonzero masses of vector fields, the superposition of the gauge triplet fields of 
$\vec{\tau}^{2}$ and of the gauge singlet field of $\tau^{4}$. Since these scalar fields do not 
couple to the lower four families (which are singlets with respect to  $\vec{\tilde{N}}_{R}$ and 
$\vec{\tilde{\tau}}^{2}$) the lower four families stay massless at this break.

At the successive break, that is at the electroweak break, several other combinations of 
$f^{\sigma}{}_{s}\, \tilde{\omega}_{ab\sigma}$, the gauge triplets of $\vec{\tilde{N}}_{L}$ and 
of $\vec{\tilde{\tau}}^{1}$ (which are orthogonal to previous triplets), together with some 
combinations of scalar fields  $f^{\sigma}{}_{s}\, \omega_{s't\sigma}$,  the gauge fields of 
$Q,\,Q'\,$  and $Y'$, gain nonzero vacuum expectation values, contributing correspondingly to 
mass matrices of the lower four families and  to  masses of the gauge fields $W^{\pm}_m$ 
and $Z_m$, influencing slightly, together  with the massive vector fields, also  mass matrices 
of the upper four families.

 Although mass matrices of the family members  are in each of the two groups of four families 
 very much related on the tree level  ($u$-quarks are related to $\nu$-leptons and $d$-quarks to 
 $e$-leptons),  the loop corrections, in which the  scalar fields of both kinds contribute, those 
 distinguishing among the families and those distinguishing  among the family members 
 ($u\,,d\,,\nu\,,e$), together with the massive vector gauge fields which distinguish only 
 among family members, start to  hopefully (as so far done calculations~\cite{albinonorma} manifest) 
 explain why are  properties of the so far observed quarks and leptons so different. Numerical 
 evaluations of the loop corrections to the tree level are in preparation (the ref.~\cite{albinonorma}).
  
 (It might be, however,  that the influence of a very special  term in higher loop corrections, 
 which influences  the neutrinos, since it transforms the right handed neutrinos into the 
 left handed  charged conjugated ones, is very strong and might be responsible for the  properties 
 of neutrinos of the lower three families.) 
 
 I simulate the {\it standard model} as the effective low energy model of the 
 {\it spin-charge-family} theory also by replacing  the operator, which in the 
 {\it spin-charge-family} theory transforms the 
 weak and hyper charges of  right handed  quarks and leptons into those of their left handed partners,  
 by a weak doublet scalar, colour singlet and of  an appropriate hyper charge. 
 While in the {\it spin-charge-family} theory the scalar dynamical fields are all doublets with respect
 to the weak charge, they all are triplets with respect to the family quantum numbers and singlets with respect 
 to three $U(1)$ charges. They  determine the Yukawa couplings. 

It is further tried to understand to 
 which extent can the scalar fields 
 originating in $\tilde{\omega}_{abs}$ and $\omega_{abs}$ spin connection dynamical fields 
 %(all in  the adjoint representations with respect to all the gauge groups) 
 be replaced by a kind 
 of a "bi-fundamental" (with respect to the family group) Yukawa scalar dynamical fields of 
 the models~\cite{belen,giudice}, in which fermion families are assumed to be members of 
 the $SU(3)$ flavour group. It seems so far that it is hard to learn something from 
 such, at least from the point of  view of the {\it spin-charge-family} theory, 
 quite artificial models designed to  
 extend further the  {\it standard model} assumption that the Higgs is in the 
 fundamental representations of the weak charge group.  The Higgs, although so far 
 very successful, seems to that it can hardly be  extended to Yukawas. 
 
 Let me repeat that the {\it spin-charge-family} theory does not  support the existence 
 of the supersymmetric partners of the so far observed fermions and gauge bosons (assuming   
 that there exist fermions with the charges in the adjoint representations and bosons 
 with the charges in the fundamental representations). 
 
 Let me add that if the {\it spin-charge-family} theory offers the right explanation for the 
 families of fermions 
 and their quantum numbers as well as for the gauge and scalar dynamical fields, then the 
 scalar dynamical fields represent new forces, as do also the Yukawas 
 of the {\it standard model}.
 
 Let me point out at the end that the {\it spin-charge-family} theory,  offering the
 explanation for the appearance of spin, charges and families of fermions, and for the 
 appearance of  gauge vector and scalar boson fields at low energy regime,  still needs 
 a careful studies, numerical ones and also proofs, to demonstrate that this is 
 the right next step beyond the {\it standard model}.

\appendix*

\section{Short presentation of the technique~\cite{norma,snmb:hn02hn03}}
\label{technique}

I make in this appendix a short review of the technique~\cite{{snmb:hn02hn03}}, 
initiated and developed by me when  proposing the 
{\it spin-charge-family} theory~\cite{norma,pikanorma,Portoroz03}  assuming that
all the internal degrees of freedom of spinors, with family quantum number included, are 
describable in the space of $d$-anticommuting (Grassmann) coordinates~\cite{snmb:hn02hn03}, if the 
dimension of ordinary space is also $d$. There are two kinds of operators in the Grasmann space, 
fulfilling the Clifford algebra which anticommute with one another. The technique  was further 
developed in the present shape together with H.B. Nielsen~\cite{snmb:hn02hn03} by identifying 
one kind of the Clifford objects with $\gamma^s$'s and another kind with  $\tilde{\gamma}^a$'s. 
In this last stage we  constructed a spinor basis as products of nilpotents and projections  formed  
as odd and even objects of $\gamma^a$'s, respectively, and  chosen to be eigenstates 
of a Cartan subalgebra of the Lorentz groups defined by $\gamma^a$'s and $\tilde{\gamma}^a$'s.   
The technique can be used to construct a spinor basis for any dimension $d$
and any signature in an easy and transparent way. Equipped with the graphic presentation of basic states,  
the technique offers an elegant way to see all the quantum numbers of states with respect to the two 
Lorentz groups, as well as transformation properties of the states under any Clifford algebra object.

The objects $\gamma^a$ and $\tilde{\gamma}^a$ have properties~\ref{snmb:tildegclifford},
\begin{eqnarray}
\label{gammatildegamma}
&& \{ \gamma^a, \gamma^b\}_{+} = 2\eta^{ab}\,, \quad\quad    
\{ \tilde{\gamma}^a, \tilde{\gamma}^b\}_{+}= 2\eta^{ab}\,, \quad,\quad
\{ \gamma^a, \tilde{\gamma}^b\}_{+} = 0\,,
\end{eqnarray}
for any $d$, even or odd.  $I$ is the unit element in the Clifford algebra.

The Clifford algebra objects $S^{ab}$ and $\tilde{S}^{ab}$ close the algebra of the Lorentz group 
\begin{eqnarray}
\label{sabtildesab}
\
S^{ab}: &=& (i/4) (\gamma^a \gamma^b - \gamma^b \gamma^a)\,, \nonumber\\
\tilde{S}^{ab}: &=& (i/4) (\tilde{\gamma}^a \tilde{\gamma}^b 
- \tilde{\gamma}^b \tilde{\gamma}^a)\,,\nonumber\\
 \{S^{ab}, \tilde{S}^{cd}\}_{-}&=& 0\,,\nonumber\\
\{S^{ab},S^{cd}\}_{-} &=& i(\eta^{ad} S^{bc} + \eta^{bc} S^{ad} - \eta^{ac} S^{bd} - \eta^{bd} S^{ac})\,,
\nonumber\\
\{\tilde{S}^{ab},\tilde{S}^{cd}\}_{-} &=& i(\eta^{ad} \tilde{S}^{bc} + \eta^{bc} \tilde{S}^{ad} 
- \eta^{ac} \tilde{S}^{bd} - \eta^{bd} \tilde{S}^{ac})\,,
\end{eqnarray}

We assume  the ``Hermiticity'' property for $\gamma^a$'s  and $\tilde{\gamma}^a$'s 
\begin{eqnarray}
\gamma^{a\dagger} = \eta^{aa} \gamma^a\,,\quad \quad \tilde{\gamma}^{a\dagger} = \eta^{aa} \tilde{\gamma}^a\,,
\label{cliffher}
\end{eqnarray}
in order that 
$\gamma^a$ and $\tilde{\gamma}^a$ are compatible with (\ref{gammatildegamma}) and formally unitary, 
i.e. $\gamma^{a \,\dagger} \,\gamma^a=I$ and $\tilde{\gamma}^{a\,\dagger} \tilde{\gamma}^a=I$.

One finds from Eq.(\ref{cliffher}) that $(S^{ab})^{\dagger} = \eta^{aa} \eta^{bb}S^{ab}$.

Recognizing from Eq.(\ref{sabtildesab})  that two Clifford algebra objects 
$S^{ab}, S^{cd}$ with all indices different commute, and equivalently for 
$\tilde{S}^{ab},\tilde{S}^{cd}$, we  select  the Cartan subalgebra of the algebra of the 
two groups, which  form  equivalent representations with respect to one another 
\begin{eqnarray}
S^{03}, S^{12}, S^{56}, \cdots, S^{d-1\; d}, \quad {\rm if } \quad d &=& 2n\ge 4,
\nonumber\\
S^{03}, S^{12}, \cdots, S^{d-2 \;d-1}, \quad {\rm if } \quad d &=& (2n +1) >4\,,
\nonumber\\
\tilde{S}^{03}, \tilde{S}^{12}, \tilde{S}^{56}, \cdots, \tilde{S}^{d-1\; d}, 
\quad {\rm if } \quad d &=& 2n\ge 4\,,
\nonumber\\
\tilde{S}^{03}, \tilde{S}^{12}, \cdots, \tilde{S}^{d-2 \;d-1}, 
\quad {\rm if } \quad d &=& (2n +1) >4\,.
\label{choicecartan}
\end{eqnarray}

The choice for  the Cartan subalgebra in $d <4$ is straightforward.
It is  useful  to define one of the Casimirs of the Lorentz group -  
the  handedness $\Gamma$ ($\{\Gamma, S^{ab}\}_- =0$) in any $d$ 
\begin{eqnarray}
\Gamma^{(d)} :&=&(i)^{d/2}\; \;\;\;\;\;\prod_a \quad (\sqrt{\eta^{aa}} \gamma^a), \quad {\rm if } \quad d = 2n, 
\nonumber\\
\Gamma^{(d)} :&=& (i)^{(d-1)/2}\; \prod_a \quad (\sqrt{\eta^{aa}} \gamma^a), \quad {\rm if } \quad d = 2n +1\,.
\label{hand}
\end{eqnarray}
One can proceed equivalently for $\tilde{\gamma}^a$'s.
We understand the product of $\gamma^a$'s in the ascending order with respect to 
the index $a$: $\gamma^0 \gamma^1\cdots \gamma^d$. 
It follows from Eq.(\ref{cliffher})
for any choice of the signature $\eta^{aa}$ that
$\Gamma^{\dagger}= \Gamma,\;
\Gamma^2 = I.$
We also find that for $d$ even the handedness  anticommutes with the Clifford algebra objects 
$\gamma^a$ ($\{\gamma^a, \Gamma \}_+ = 0$) , while for $d$ odd it commutes with  
$\gamma^a$ ($\{\gamma^a, \Gamma \}_- = 0$). 

To make the technique simple we introduce the graphic presentation 
as follows (Eq.~(\ref{snmb:eigensab}))
\begin{eqnarray}
\stackrel{ab}{(k)}:&=& 
\frac{1}{2}(\gamma^a + \frac{\eta^{aa}}{ik} \gamma^b)\,,\quad \quad
\stackrel{ab}{[k]}:=
\frac{1}{2}(1+ \frac{i}{k} \gamma^a \gamma^b)\,,\nonumber\\
\stackrel{+}{\circ}:&=& \frac{1}{2} (1+\Gamma)\,,\quad \quad
\stackrel{-}{\bullet}:= \frac{1}{2}(1-\Gamma),
\label{signature}
\end{eqnarray}
where $k^2 = \eta^{aa} \eta^{bb}$.
One can easily check by taking into account the Clifford algebra relation 
(Eq.\ref{gammatildegamma}) and the
definition of $S^{ab}$ and $\tilde{S}^{ab}$ (Eq.\ref{sabtildesab})
that if one multiplies from the left hand side by $S^{ab}$ or $\tilde{S}^{ab}$ the Clifford 
algebra objects $\stackrel{ab}{(k)}$
and $\stackrel{ab}{[k]}$,
it follows that
\begin{eqnarray}
        S^{ab}\, \stackrel{ab}{(k)}= \frac{1}{2}\,k\, \stackrel{ab}{(k)}\,,\quad \quad 
        S^{ab}\, \stackrel{ab}{[k]}= \frac{1}{2}\,k \,\stackrel{ab}{[k]}\,,\nonumber\\
\tilde{S}^{ab}\, \stackrel{ab}{(k)}= \frac{1}{2}\,k \,\stackrel{ab}{(k)}\,,\quad \quad 
\tilde{S}^{ab}\, \stackrel{ab}{[k]}=-\frac{1}{2}\,k \,\stackrel{ab}{[k]}\,,
\label{grapheigen}
\end{eqnarray}
which means that we get the same objects back multiplied by the constant $\frac{1}{2}k$ in the case 
of $S^{ab}$, while $\tilde{S}^{ab}$ multiply $\stackrel{ab}{(k)}$ by $k$ and $\stackrel{ab}{[k]}$ 
by $(-k)$ rather than $(k)$. 
%%%%%%%%%%%%
This also means that when 
$\stackrel{ab}{(k)}$ and $\stackrel{ab}{[k]}$ act from the left hand side on  a
vacuum state $|\psi_0\rangle$ the obtained states are the eigenvectors of $S^{ab}$.
We further recognize~(Eq.~\ref{snmb:graphgammaaction},\ref{snmb:gammatilde}) that $\gamma^a$ 
transform  $\stackrel{ab}{(k)}$ into  $\stackrel{ab}{[-k]}$, never to $\stackrel{ab}{[k]}$, 
while $\tilde{\gamma}^a$ transform  $\stackrel{ab}{(k)}$ into $\stackrel{ab}{[k]}$, never to 
$\stackrel{ab}{[-k]}$ 
\begin{eqnarray}
%\label{snmb:graphgammatilgegammaaction}
&&\gamma^a \stackrel{ab}{(k)}= \eta^{aa}\stackrel{ab}{[-k]},\; 
\gamma^b \stackrel{ab}{(k)}= -ik \stackrel{ab}{[-k]}, \; 
\gamma^a \stackrel{ab}{[k]}= \stackrel{ab}{(-k)},\; 
\gamma^b \stackrel{ab}{[k]}= -ik \eta^{aa} \stackrel{ab}{(-k)}\,,\nonumber\\
&&\tilde{\gamma^a} \stackrel{ab}{(k)} = - i\eta^{aa}\stackrel{ab}{[k]},\;
\tilde{\gamma^b} \stackrel{ab}{(k)} =  - k \stackrel{ab}{[k]}, \;
\tilde{\gamma^a} \stackrel{ab}{[k]} =  \;\;i\stackrel{ab}{(k)},\; 
\tilde{\gamma^b} \stackrel{ab}{[k]} =  -k \eta^{aa} \stackrel{ab}{(k)}\,. 
\label{snmb:gammatildegamma}
\end{eqnarray}
>From Eq.(\ref{snmb:gammatildegamma}) it follows
\begin{eqnarray}
\label{stildestrans}
S^{ac}\stackrel{ab}{(k)}\stackrel{cd}{(k)} &=& -\frac{i}{2} \eta^{aa} \eta^{cc} 
\stackrel{ab}{[-k]}\stackrel{cd}{[-k]}\,,\,\quad\quad
\tilde{S}^{ac}\stackrel{ab}{(k)}\stackrel{cd}{(k)} = \frac{i}{2} \eta^{aa} \eta^{cc} 
\stackrel{ab}{[k]}\stackrel{cd}{[k]}\,,\,\nonumber\\
S^{ac}\stackrel{ab}{[k]}\stackrel{cd}{[k]} &=& \frac{i}{2}  
\stackrel{ab}{(-k)}\stackrel{cd}{(-k)}\,,\,\quad\quad
\tilde{S}^{ac}\stackrel{ab}{[k]}\stackrel{cd}{[k]} = -\frac{i}{2}  
\stackrel{ab}{(k)}\stackrel{cd}{(k)}\,,\,\nonumber\\
S^{ac}\stackrel{ab}{(k)}\stackrel{cd}{[k]}  &=& -\frac{i}{2} \eta^{aa}  
\stackrel{ab}{[-k]}\stackrel{cd}{(-k)}\,,\,\quad\quad
\tilde{S}^{ac}\stackrel{ab}{(k)}\stackrel{cd}{[k]} = -\frac{i}{2} \eta^{aa}  
\stackrel{ab}{[k]}\stackrel{cd}{(k)}\,,\,\nonumber\\
S^{ac}\stackrel{ab}{[k]}\stackrel{cd}{(k)} &=& \frac{i}{2} \eta^{cc}  
\stackrel{ab}{(-k)}\stackrel{cd}{[-k]}\,,\,\quad\quad
\tilde{S}^{ac}\stackrel{ab}{[k]}\stackrel{cd}{(k)} = \frac{i}{2} \eta^{cc}  
\stackrel{ab}{(k)}\stackrel{cd}{[k]}\,. 
\end{eqnarray}
>From Eqs.~(\ref{stildestrans}) we conclude that $\tilde{S}^{ab}$ generate the 
equivalent representations with respect to $S^{ab}$ and opposite. 

Let us deduce some useful relations

\begin{eqnarray}
\stackrel{ab}{(k)}\stackrel{ab}{(k)}& =& 0\,, \quad \quad \stackrel{ab}{(k)}\stackrel{ab}{(-k)}
= \eta^{aa}  \stackrel{ab}{[k]}\,, \quad \stackrel{ab}{(-k)}\stackrel{ab}{(k)}=
\eta^{aa}   \stackrel{ab}{[-k]}\,,\quad
\stackrel{ab}{(-k)} \stackrel{ab}{(-k)} = 0\,, \nonumber\\
\stackrel{ab}{[k]}\stackrel{ab}{[k]}& =& \stackrel{ab}{[k]}\,, \quad \quad
\stackrel{ab}{[k]}\stackrel{ab}{[-k]}= 0\,, \;\;\quad \quad  \quad \stackrel{ab}{[-k]}\stackrel{ab}{[k]}=0\,,
 \;\;\quad \quad \quad \quad \stackrel{ab}{[-k]}\stackrel{ab}{[-k]} = \stackrel{ab}{[-k]}\,,
 \nonumber\\
\stackrel{ab}{(k)}\stackrel{ab}{[k]}& =& 0\,,\quad \quad \quad \stackrel{ab}{[k]}\stackrel{ab}{(k)}
=  \stackrel{ab}{(k)}\,, \quad \quad \quad \stackrel{ab}{(-k)}\stackrel{ab}{[k]}=
 \stackrel{ab}{(-k)}\,,\quad \quad \quad 
\stackrel{ab}{(-k)}\stackrel{ab}{[-k]} = 0\,,
\nonumber\\
\stackrel{ab}{(k)}\stackrel{ab}{[-k]}& =&  \stackrel{ab}{(k)}\,,
\quad \quad \stackrel{ab}{[k]}\stackrel{ab}{(-k)} =0,  \quad \quad 
\quad \stackrel{ab}{[-k]}\stackrel{ab}{(k)}= 0\,, \quad \quad \quad \quad
\stackrel{ab}{[-k]}\stackrel{ab}{(-k)} = \stackrel{ab}{(-k)}.
\label{graphbinoms}
\end{eqnarray}
We recognize in  the first equation of the first row and the first equation of the second row
the demonstration of the nilpotent and the projector character of the Clifford algebra objects 
$\stackrel{ab}{(k)}$ and $\stackrel{ab}{[k]}$, respectively. 
Defining
\begin{eqnarray}
\stackrel{ab}{\tilde{(\pm i)}} = 
\frac{1}{2} \, (\tilde{\gamma}^a \mp \tilde{\gamma}^b)\,, \quad
\stackrel{ab}{\tilde{(\pm 1)}} = 
\frac{1}{2} \, (\tilde{\gamma}^a \pm i\tilde{\gamma}^b)\,, 
%\stackrel{ab}{\tilde{[\pm i]}} = \frac{1}{2} (1 \pm \tilde{\gamma}^a \tilde{\gamma}^b), \quad
%\stackrel{ab}{\tilde{[\pm 1]}} = \frac{1}{2} (1 \pm i \tilde{\gamma}^a \tilde{\gamma}^b). \nonumber
\label{deftildefun}
\end{eqnarray}
one recognizes that
\begin{eqnarray}
\stackrel{ab}{\tilde{( k)}} \, \stackrel{ab}{(k)}& =& 0\,, 
\quad \;
\stackrel{ab}{\tilde{(-k)}} \, \stackrel{ab}{(k)} = -i \eta^{aa}\,  \stackrel{ab}{[k]}\,,
\quad\;
\stackrel{ab}{\tilde{( k)}} \, \stackrel{ab}{[k]} = i\, \stackrel{ab}{(k)}\,,
\quad\;
\stackrel{ab}{\tilde{( k)}}\, \stackrel{ab}{[-k]} = 0\,.
\label{graphbinomsfamilies}
\end{eqnarray}
Recognizing that
\begin{eqnarray}
\stackrel{ab}{(k)}^{\dagger}=\eta^{aa}\stackrel{ab}{(-k)}\,,\quad
\stackrel{ab}{[k]}^{\dagger}= \stackrel{ab}{[k]}\,,
\label{graphherstr}
\end{eqnarray}
we define a vacuum state $|\psi_0>$ so that one finds
\begin{eqnarray}
< \;\stackrel{ab}{(k)}^{\dagger}
 \stackrel{ab}{(k)}\; > = 1\,, \nonumber\\
< \;\stackrel{ab}{[k]}^{\dagger}
 \stackrel{ab}{[k]}\; > = 1\,.
\label{graphherscal}
\end{eqnarray}

Taking into account the above equations it is easy to find a Weyl spinor irreducible representation
for $d$-dimensional space, with $d$ even or odd.

For $d$ even we simply make a starting state as a product of $d/2$, let us say, only nilpotents 
$\stackrel{ab}{(k)}$, one for each $S^{ab}$ of the Cartan subalgebra  elements (Eq.(\ref{choicecartan})),  
applying it on an (unimportant) vacuum state. 
For $d$ odd the basic states are products
of $(d-1)/2$ nilpotents and a factor $(1\pm \Gamma)$.  
Then the generators $S^{ab}$, which do not belong 
to the Cartan subalgebra, being applied on the starting state from the left, 
 generate all the members of one
Weyl spinor.  
\begin{eqnarray}
\stackrel{0d}{(k_{0d})} \stackrel{12}{(k_{12})} \stackrel{35}{(k_{35})}\cdots \stackrel{d-1\;d-2}{(k_{d-1\;d-2})}
\psi_0 \nonumber\\
\stackrel{0d}{[-k_{0d}]} \stackrel{12}{[-k_{12}]} \stackrel{35}{(k_{35})}\cdots \stackrel{d-1\;d-2}{(k_{d-1\;d-2})}
\psi_0 \nonumber\\
\stackrel{0d}{[-k_{0d}]} \stackrel{12}{(k_{12})} \stackrel{35}{[-k_{35}]}\cdots \stackrel{d-1\;d-2}{(k_{d-1\;d-2})}
\psi_0 \nonumber\\
\vdots \nonumber\\
\stackrel{0d}{[-k_{0d}]} \stackrel{12}{(k_{12})} \stackrel{35}{(k_{35})}\cdots \stackrel{d-1\;d-2}{[-k_{d-1\;d-2}]}
\psi_0 \nonumber\\
\stackrel{od}{(k_{0d})} \stackrel{12}{[-k_{12}]} \stackrel{35}{[-k_{35}]}\cdots \stackrel{d-1\;d-2}{(k_{d-1\;d-2})}
\psi_0 \nonumber\\
\vdots 
\label{graphicd}
\end{eqnarray}
All the states have the handedness $\Gamma $, since $\{ \Gamma, S^{ab}\} = 0$. 
States, belonging to one multiplet  with respect to the group $SO(q,d-q)$, that is to one
irreducible representation of spinors (one Weyl spinor), can have any phase. We made a choice
of the simplest one, taking all  phases equal to one.

The above graphic representation demonstrate that for $d$ even 
all the states of one irreducible Weyl representation of a definite handedness follow from a starting state, 
which is, for example, a product of nilpotents $\stackrel{ab}{(k_{ab})}$, by transforming all possible pairs
of $\stackrel{ab}{(k_{ab})} \stackrel{mn}{(k_{mn})}$ into $\stackrel{ab}{[-k_{ab}]} \stackrel{mn}{[-k_{mn}]}$.
There are $S^{am}, S^{an}, S^{bm}, S^{bn}$, which do this.
The procedure gives $2^{(d/2-1)}$ states. A Clifford algebra object $\gamma^a$ being applied from the left hand side,
transforms  a 
Weyl spinor of one handedness into a Weyl spinor of the opposite handedness. Both Weyl spinors form a Dirac 
spinor.

For $d$ odd a Weyl spinor has besides a product of $(d-1)/2$ nilpotents or projectors also either the
factor $\stackrel{+}{\circ}:= \frac{1}{2} (1+\Gamma)$ or the factor
$\stackrel{-}{\bullet}:= \frac{1}{2}(1-\Gamma)$.  
As in the case of $d$ even, all the states of one irreducible 
Weyl representation of a definite handedness follow from a starting state, 
which is, for example, a product of $(1 + \Gamma)$ and $(d-1)/2$ nilpotents $\stackrel{ab}{(k_{ab})}$, by 
transforming all possible pairs
of $\stackrel{ab}{(k_{ab})} \stackrel{mn}{(k_{mn})}$ into $\stackrel{ab}{[-k_{ab}]} \stackrel{mn}{[-k_{mn}]}$.
But $\gamma^a$'s, being applied from the left hand side, do not change the handedness of the Weyl spinor, 
since $\{ \Gamma,
\gamma^a \}_- =0$ for $d$ odd. 
A Dirac and a Weyl spinor are for $d$ odd identical and a ''family'' 
has accordingly $2^{(d-1)/2}$ members of basic states of a definite handedness.

We shall speak about left handedness when $\Gamma= -1$ and about right
handedness when $\Gamma =1$ for either $d$ even or odd.

While $S^{ab}$ which do not belong to the Cartan subalgebra (Eq.~(\ref{choicecartan})) generate 
all the states of one representation, generate $\tilde{S}^{ab}$ which do not belong to the 
Cartan subalgebra(Eq.~(\ref{choicecartan})) the states of $2^{d/2-1}$ equivalent representations.

Making a choice of the Cartan subalgebra set of the algebra $S^{ab}$ and $\tilde{S}^{ab}$  
\begin{eqnarray}
S^{03}, S^{12}, S^{56}, S^{78}, S^{9 \;10}, S^{11\;12}, S^{13\; 14}\,,\nonumber\\
\tilde{S}^{03}, \tilde{S}^{12}, \tilde{S}^{56}, \tilde{S}^{78}, \tilde{S}^{9 \;10}, 
\tilde{S}^{11\;12}, \tilde{S}^{13\; 14}\,,
\label{cartan}
\end{eqnarray}
a left handed ($\Gamma^{(1,13)} =-1$) eigen state of all the members of the 
Cartan  subalgebra, representing a weak chargeless  $u_{R}$-quark with spin up, hypercharge ($2/3$) 
and  colour ($1/2\,,1/(2\sqrt{3})$), for example, can be written as %(Eq.(\ref{cartan})) 
\begin{eqnarray}
&& \stackrel{03}{(+i)}\stackrel{12}{(+)}|\stackrel{56}{(+)}\stackrel{78}{(+)}
||\stackrel{9 \;10}{(+)}\stackrel{11\;12}{(-)}\stackrel{13\;14}{(-)} |\psi \rangle = \nonumber\\
&&\frac{1}{2^7} 
(\gamma^0 -\gamma^3)(\gamma^1 +i \gamma^2)| (\gamma^5 + i\gamma^6)(\gamma^7 +i \gamma^8)||
\nonumber\\
&& (\gamma^9 +i\gamma^{10})(\gamma^{11} -i \gamma^{12})(\gamma^{13}-i\gamma^{14})
|\psi \rangle \,.
\label{start}
\end{eqnarray}
This state is an eigenstate of all $S^{ab}$ and $\tilde{S}^{ab}$ which are members of the Cartan 
subalgebra (Eq.~(\ref{cartan})). 

The operators $ \tilde{S}^{ab}$, which do not belong to the Cartan subalgebra (Eq.~(\ref{cartan})),  
generate families from the starting $u_R$ quark, transforming $u_R$ quark from Eq.~(\ref{start}) 
to the $u_R$ of another family,  keeping all the properties with respect to $S^{ab}$ unchanged.
In particular $\tilde{S}^{01}$ applied on a right handed $u_R$-quark, weak chargeless,  with spin up,
hypercharge ($2/3$) and the colour charge ($1/2\,,1/(2\sqrt{3})$) from Eq.~(\ref{start}) generates a 
state which is again  a right handed $u_{R}$-quark,  weak chargeless,  with spin up,
hypercharge ($2/3$)
and the colour charge ($1/2\,,1/(2\sqrt{3})$)
\begin{eqnarray}
\label{tildesabfam}
\tilde{S}^{01}\;
\stackrel{03}{(+i)}\stackrel{12}{(+)}| \stackrel{56}{(+)} \stackrel{78}{(+)}||
\stackrel{9 10}{(+)} \stackrel{11 12}{(-)} \stackrel{13 14}{(-)}= -\frac{i}{2}\,
&&\stackrel{03}{[\,+i]} \stackrel{12}{[\,+\,]}| \stackrel{56}{(+)} \stackrel{78}{(+)}||
\stackrel{9 10}{(+)} \stackrel{11 12}{(-)} \stackrel{13 14}{(-)}\,.
\end{eqnarray}

Below some useful relations~\cite{pikanorma} are presented 
\begin{eqnarray}
\label{plusminus}
N^{\pm}_{+}         &=& N^{1}_{+} \pm i \,N^{2}_{+} = 
 - \stackrel{03}{(\mp i)} \stackrel{12}{(\pm )}\,, \quad N^{\pm}_{-}= N^{1}_{-} \pm i\,N^{2}_{-} = 
  \stackrel{03}{(\pm i)} \stackrel{12}{(\pm )}\,,\nonumber\\
\tilde{N}^{\pm}_{+} &=& - \stackrel{03}{\tilde{(\mp i)}} \stackrel{12}{\tilde{(\pm )}}\,, \quad 
\tilde{N}^{\pm}_{-}= %\tilde{N}^{1}_{-} \pm i\,\tilde{N}^{2}_{-} = 
  \stackrel{03} {\tilde{(\pm i)}} \stackrel{12} {\tilde{(\pm )}}\,,\nonumber\\ 
\tau^{1\pm}         &=& (\mp)\, \stackrel{56}{(\pm )} \stackrel{78}{(\mp )} \,, \quad   
\tau^{2\mp}=            (\mp)\, \stackrel{56}{(\mp )} \stackrel{78}{(\mp )} \,,\nonumber\\ 
\tilde{\tau}^{1\pm} &=& (\mp)\, \stackrel{56}{\tilde{(\pm )}} \stackrel{78}{\tilde{(\mp )}}\,,\quad   
\tilde{\tau}^{2\mp}= (\mp)\, \stackrel{56}{\tilde{(\mp )}} \stackrel{78}{\tilde{(\mp )}}\,.
\end{eqnarray}

\end{document}